\documentclass[letterpaper]{article}
\usepackage{jheppub}
\pdfoutput=1
\usepackage{graphicx, caption, subcaption}  
\usepackage{dcolumn}   
\usepackage{amsmath, amssymb}        
\usepackage{color,latexsym, array,multirow,  verbatim, enumerate}
\usepackage{url}
\graphicspath{ {/Users/tuhin/Projects/vvresonance/Plotting/} }

\newcommand{\tev}{~\text{TeV}}
\newcommand{\gev}{~\text{GeV}}

\newcommand{\ifb}{~\text{fb}^{-1}}

\title{A Cautionary Tale of Mis-measured Tails from $q/g$ Bias}
\author[a]{Adam Martin,} 
\author[b,c]{and Tuhin S.~Roy}
\affiliation[a]{Department of Physics, University of Notre Dame, Notre Dame, IN 46556, USA}

\affiliation[b]{Department of Theoretical Physics, Tata Institute of Fundamental Research, Mumbai 400005, India}
\affiliation[c]{Theory Division T-2, Los Alamos National laboratory, Los Alamos, NM 87545, USA}

\date{\today}
\abstract{ Jet substructure techniques such as subjet $p_T$-asymmetry, mass-drop, and grooming have become powerful and widely used tools in experimental searches at the LHC. While these tools provide much-desired handles to separate signal from background, they can introduce unexpected mass scales into the analysis. These scales may be misinterpreted as excesses if these are not correctly incorporated into background modeling. As an example, we study the ATLAS hadronic di-$W/Z$ resonance search. There, we find that the substructure analysis -- in particular the combination of a subjet asymmetry cut with the requirement on the number of tracks within a jet -- induces a mass scale where the dominant partonic subprocess in the background changes from $pp \to g \!+\! q/\bar q$ to $pp \to q\bar{q}$.  In light of this scale, modeling the QCD background using a simple smooth function with monotonically decreasing slope appears insufficient.}

\preprint{}

\begin{document}
\maketitle

\flushbottom

\section{\label{sec:1}Introduction}

The Standard Model of particle physics has long provided us with guidance towards what new phenomena to expect and how to find new particles. Now that the Higgs boson has been discovered, this guidance is gone. In its place, experimental searches are often inspired by various models of physics Beyond the Standard Model (BSM). While it is true that signals of BSM models such as supersymmetry, extra dimensions, or strong dynamics are often characterized by widely diverse  configurations in the final state particles and often with varied topologies, it has also been internalized not to design searches based \textit{only} on `well-motivated' BSM scenarios and to perform general purpose searches as well. From the experimental side, the arguments are straightforward: searches should be exhaustive in relation to what the designs and the performances of the colliders and the detectors can deliver, and theorists' prejudices should be a secondary concern. 

One example of a general purpose search is the search for dijet resonances. One searches for a bump in the falling continuum of the invariant mass of two jets observed in events consisting of say, exactly two jets, irrespective of whether or not a given BSM model has already ruled out the existence of such a particle based on some other search. While general purpose searches cover a large class of potential BSM scenarios, they usually have fewer handles to distinguish signal from background than a search dedicated to a particular model. For the case of dijet resonance searches, one traditionally only has the jet energies, angular distribution, and dijet mass as handles. However, if we narrow our search to resonances that decay to a pair of massive, hadronically decaying particles, we can bring the tools of jet substructure to bear, thus gaining ways to distinguish signal from background. Substructure techniques, proposed as early as in Ref.~\cite{Seymour:1993mx}, have been tuned and improved over the years: to increase tagging efficiencies of jets arising from the decay of boosted heavy particles and even of standard  detected objects such as leptons, photons, heavy flavor jets etc.~\cite{Seymour:1993mx, Butterworth:2007ke, Brooijmans:1077731,Butterworth:2008iy, Butterworth:2009qa, Kaplan:2008ie,Thaler:2008ju,Almeida:2008yp,Plehn:2009rk, Thaler:2010tr, Soper:2011cr, Englert:2011iz, Thaler:2011gf, Plehn:2011tg,Ellis:2012sd, Ellis:2012zp, Soper:2012pb,Ortiz:2014iza, Ellis:2012sn, Pedersen:2015knf, Pedersen:2015hka}; to measure properties of jets~\cite{Gallicchio:2010sw, Gallicchio:2011xc, Gallicchio:2011xq,Gallicchio:2012ez, Krohn:2012fg}; and to  remove unwanted radiation from jets (namely, to groom  jets) associated with any event in a hadron collider~\cite{Butterworth:2008iy, Butterworth:2008sd, Butterworth:2008tr, Ellis:2009su, Ellis:2009me,Krohn:2009th, Soyez:2012hv, Cacciari:2014jta,Cacciari:2014gra, Bertolini:2014bba}. The goal of this paper is not to add to this already impressive list of tools, but to urge more caution while using these tools. In particular, our purpose is to point out that substructure-based analyses may introduce unexpected scales in the background (often due to QCD), which can give rise to miscalculated distributions and false excesses. We do not mean to imply spurious scales are introduced only by substructure cuts, as kinematic cuts ($p_T, \eta_j$, etc.) certainly implant scales into the background. Our point is rather that {\em all} scales need to be correctly incorporated into the background model.

In this paper, we use the recent ATLAS~\cite{Aad:2015owa} analysis as a case study to illustrate the above point. The ATLAS search was designed to find heavy and narrow resonances decaying to $WW$, $ZZ$, or  $WZ$. For resonances heavier than $1\, \tev$, the target region of the study, the daughter $W/Z$ have such high transverse momenta ($p_T$) that their subsequent decay products are nearly collimated. The search, therefore, became a search for dijet events with each jet containing all decay products of a $W/Z$. Naively, one might expect that forcing the mass of each jet to lie in the $W/Z$ mass window is an effective way to separate signal from background. Unfortunately, this does not work well. First, for the range of jet $p_T$ in this study, a large number of background (QCD) jets also have a mass in the $W/Z$ window.  Second, noise due to initial state radiation (ISR), multiple interactions, and pile-up all contribute and  make jets more massive. As the jet mass distribution in QCD is given by a falling function (in the range of interests), effectively more and more jets move into the signal window due to noise.

The ATLAS collaboration uses three techniques of substructure physics to reduce the background. The first is the application of the idea proposed in Ref.~\cite{Butterworth:2008iy}, where they implement the so-called mass-drop + asymmetry cuts, which distinguish jets containing massive particle decay products from jets due to QCD. The second procedure (named ``filtering"), also proposed in Ref.~\cite{Butterworth:2008iy}, grooms the jet to remove elements due to noise. Thirdly, they count the number of charged tracks (say, $n_\text{track}$) associated with ungroomed jets and get rid of jets with a large number of tracks. The track count is a well tested measure to discriminate gluon-initiated jets from quark-initiated jets or, in this case, di-quark (from $W/Z$ decay)-initiated jets.  Since a significant part of the background contains gluon-initiated jets, one again expects a good reduction of the background.  Combining the tools mentioned above with conventional cuts (such as a cut on the angle between the jets, etc.), ATLAS extracted impressive separation of signal and background. In fact, they reported an excess of events (a bump-like feature on top of the background) between ($1.7\,\tev$ - $2.2\,\tev$) after analyzing $20\ifb$ of data from $8\,\tev$ collisions. 

Not surprisingly, the ATLAS report was followed by a rush of papers which tried to explain the excess with new physics models (see Ref.~\cite{Brehmer:2015dan} and references within), and little effort was spared in order to comprehend the analysis critically. A special mention is Ref.~\cite{Goncalves:2015yua}, where a clear, systematic study of the analysis was provided. The authors criticized many aspects of the parameters used in the substructure analysis and also laid out clearly the scope for improvements. This paper continues in the steps of~\cite{Goncalves:2015yua} and questions the validity of the ATLAS background model, taken to be a smoothly falling function.  Such an approach makes sense when one does not expect any specific scale appearing in the background. We actually find results contrary to the claim. To be specific, consider the dijet background due to QCD. Before any substructure variables are introduced, the events are dominated by jets initiated by gluons. The substructure variables generically (and the cut on the number of tracks in a jet in particular) bring down the fractions of gluon jets with respect to jets initiated by quarks. Depending on the exact values of the cuts, we find that a scale arises in the dijet-mass spectrum, below which the background is dominated by $g \!+\! q/\bar q$-type events (meaning, $p + p \to g +  q/\bar q$ at parton level) and above which $q\,\bar q$-type events take over.  Both subprocess ($g \!+\! q/\bar q$ and $q\,\bar q$) are characterized by smoothly falling distributions, but the slope is different between the two. Thus, once the subprocesses are combined, the dijet mass spectrum ends up with a feature at the transition point that deviates from a single, smoothly falling distribution. When viewed with limited statistics -- as in the ATLAS analysis where the tail is populated by $O(20)$ events, this feature can mimic a bump-like feature. 

In the mass drop + asymmetry + $n_\text{track }$ cut analysis, we find that the crossover scale depends critically on the $n_\text{track}$ cut. We also explore how the crossover scale changes under a relative quark jet vs. gluon jet mismeasurement and the collider center of mass (c.o.m.) energy. Our motivation for introducing a mismeasurement is that, while detector simulation programs include rough jet resolution, the schemes employed are driven by the gross properties of jets (energy, angle, etc.) and may be insufficient for detailed substructure variables.  Also, given that the inability of Monte Carlo programs to adequately describe the different detector response to quark vs. gluon jets has been used in the past to explain excesses -- most notably the $W+jj$ excess observed by CDF in 2011~\cite{Aaltonen:2014mdq,Aaltonen:2011mk} -- it is worth investigating the robustness of the ATLAS analysis in the presence of slight relative $q/g$ mismeasurement.

We emphasize that even though we use  the  ATLAS report as a case study to illustrate that a more careful understanding of the background is warranted when one uses substructure variables, the scope of this work is more general and applies to other jet substructure searches. In particular, we note that a similar physics signal has also been studied by the CMS collaboration with $8\tev$~\cite{Khachatryan:2014hpa},  and $13\tev$~\cite{CMS-PAS-EXO-15-002} data,  as well as by ATLAS using $13\tev$~\cite{ATLAS-CONF-2015-073} data. The results stated in this paper are relevant for all these analyses. However, each of these studies are qualitatively different, and, as a consequence the magnitude of the effect stated in this paper will be quantitatively different for each of these cases.  Analyzing every one of these studies is beyond the scope of this paper, and we will stick with the analysis as reported in Ref.~\cite{Aad:2015owa}.

The rest of this paper is organized as follows: in Section.~\ref{sec:2} we discuss how various substructure-based observables alter the quark/gluon content of events due to QCD; in Section.~\ref{sec:3} we demonstrate that the ATLAS analysis of Ref.~\cite{Aad:2015owa}, in particular, can give rise to a bump like feature in events due to QCD, at a scale generated by the use of various substructure based cuts as well as on relative $q/g$ energy mis-measurements; and finally  in Section.~\ref{sec:4} we conclude. 

\section{\label{sec:2}Quark vs. Gluon Bias from   Substructure Analyses}

It is well appreciated and understood that jet substructure variables can play crucial roles in reducing the backgrounds due to QCD. From discovering new physics~\cite{Butterworth:2008iy, Kribs:2009yh, Kribs:2010hp, Kribs:2010ii, Englert:2011iz} to measuring cross-sections~\cite{Ellis:2014eya}  of various standard model processes, these variables have been shown  to be useful both by experimentalists~\cite{Altheimer:2013yza} and theorists. The purpose of this section is to demonstrate that, in addition to reducing the background, on applying these variables one inadvertently also ends up changing the nature of the background. Let us be more precise.  Various grooming algorithms such as filtering~\cite{Butterworth:2008iy, Butterworth:2008sd, Butterworth:2008tr}, trimming~\cite{Krohn:2009th}, etc., reduce the bin-by-bin count in jet-mass distribution for large jet-masses  when applied to QCD jets. In this section we show that after these techniques are used, the quark-gluon fractions in each bin is also altered, \textit{i.e.}, bins originally occupied mostly with gluon-initiated jets may get flooded by quark-initiated jets. 

We begin with a sample of Cambridge-Aachen  (C/A)~\cite{Dokshitzer:1997in, Wobisch:1998wt, Wobisch:2000dk} jets of $R=1.2$,  constructed out of  QCD dijet-events  (details of the simulation will be given in Section~\ref{sec:3}). We split the sample based on partons initiating the jets. The jet mass and $p_T$ distributions of the samples are shown in Fig.~\ref{fig:1}. The QCD events are made with $\hat{p}_T > 500\gev$\footnote{Here the hat denotes a parton-level variable.}, whereas the jets are constructed with $p_T > 550\gev$. As a result, the $p_T$ spectra obtain a peak like feature. Even though the gluon and the quark initiated jets have similar $p_T$ distributions, in general, gluon-initiated jets obtain more masses since these have larger probabilities for energetic and large angle emissions.  Further, since we use jets with large area,  all jets accumulate  a large amount of noise. This shifts the mass spectra for both kinds of jets to higher values. 
\begin{figure}[h]
    \centering
    \includegraphics[width=\textwidth]{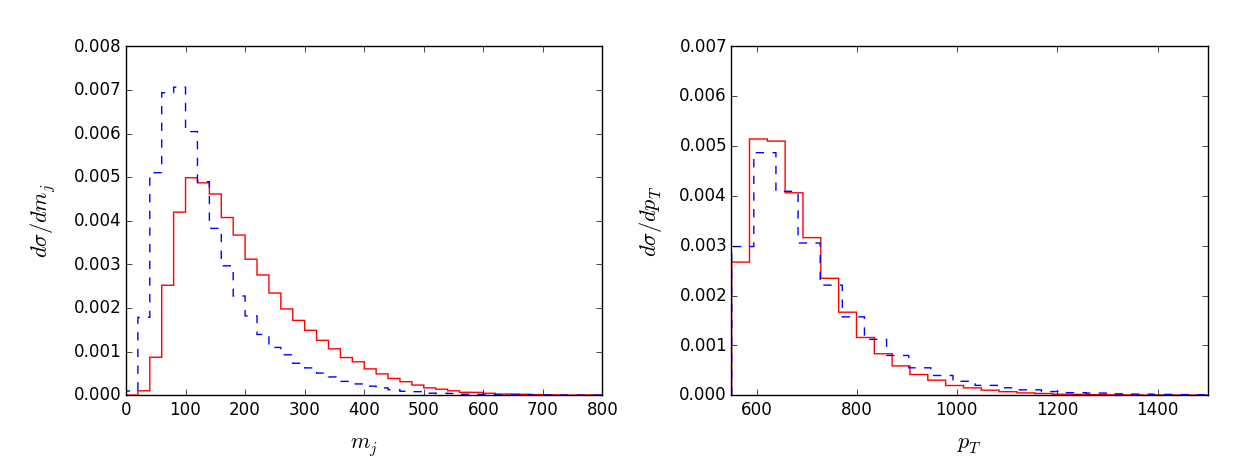}
    \caption{\label{fig:1} The mass and $p_T$ distribution of the jet sample before filtering. The  quark and gluon initiated jets are represented by  blue (dashed) and  red (solid) lines respectively.}
\end{figure}

In order to understand the effect of grooming on these jets, we subject them to filtering and trimming. While these processes sound similar, their effect on jet masses can be dramatically different when considered on a jet-by-jet basis. Both filtering and trimming involve re-clustering the constituents of a jet with a smaller radius (denoted here by  $R=R_\text{filter} $ and  $R=R_\text{trim} $ for filtering and trimming respectively). In the case of filtering, a fixed number of hardest subjets (namely, $n_\text{filter}$) are kept, whereas in trimming all subjets with $p_T > f_{\text{trim}} \: p_{T_j}$ are kept. In this section (and throughout this paper) we use the standard parameters for filtering and trimming, namely:
\begin{equation}
	R_\text{filter} \ = \ 0.3 \, , \qquad n_\text{filter} \ = \ 3 \, ; \qquad \text{and} \qquad 
	R_\text{trim} \ = \ 0.2 \, , \qquad f_\text{trim} \ = \ 0.03 \, .
\label{eq:groomparam}	
\end{equation}
We use C/A algorithm to re-cluster jets in case of filtering, whereas we use the $k_T$-algorithm~\cite{Ellis:1993tq,Catani:1993hr} for trimming as recommended by the authors.

The results are presented in Figs.~\ref{fig:2} and \ref{fig:3}. We quantify the degree of grooming as $m^\text{filtered}_j/m_j$ and $m^\text{trimmed}_j/m_j$,  where $m_j$ represents the ungroomed jet-mass,   $m^\text{filtered}_j$  and $m^\text{trimmed}_j$  represent groomed jet-masses  after the jet goes through filtering and trimming respectively. Note that, for a given jet a quantity of interest is $m_j/p_{T_j}$, which gives the angular size of the jet. In Figs.~\ref{fig:2} and \ref{fig:3} we have plotted the probability density functions (pdfs) for both gluon and quark initiated jets as functions of the degree of grooming and $m_j/p_{T_j}$.  There are two lessons: $(i.)$  the figures demonstrate that both the grooming algorithms treat jets differently based on the partons initiating the jets, and $(ii.)$  this $q/g$ discrimination depends sensitively on  the grooming algorithm.  
\begin{figure}[h]
    \centering
    \includegraphics[width=\textwidth]{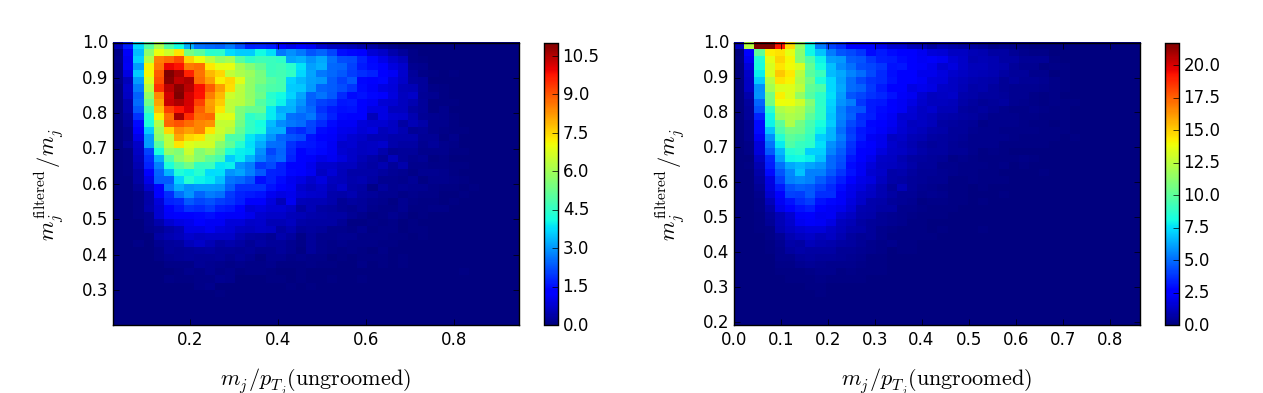}
    \caption{\label{fig:2}The effect of filtering on jet mass distributions on the gluon (left) and the quark (right) initiated jets. The pdfs in each cases are plotted as functions of  $m_j/p_T$ of the ungroomed jets and the mass fraction $m^\text{filtered}_j/m_j$.}
\end{figure}

In case of filtering, the gluon initiated jets are groomed significantly more than the quark initiated jets with the same angular dimensions.  Understanding this behavior is straightforward and has to do with the multiplicities of particles in a jet. Note that even though gluons and quarks differ both in spins and color charges, the difference in multiplicities of particles in jets initiated by gluons and quarks is mostly due to their color charges. In fact, at leading order, the multiplicity of any type of particle in gluon-initiated jets  is enhanced w.r.t. the quark initiated jets by simply the group theory factors (namely, $C_A/C_F$)~\cite{Brodsky:1976mg}. The energy dependence of this factor of enhancement arises at NLO via $\alpha_s$. Note that  a significant amount of theoretical effort has gone towards understanding the ratio of average multiplicities in quark vs gluon initiated jets (denoted by $\langle N_g \rangle$ and $\langle N_q \rangle$ respectively). At NNLO, for example, it was shown in Ref.~\cite{Gaffney:1984yd} that  
\begin{equation}
\frac{\langle N_g \rangle}{\langle N_q \rangle} \ = \ \frac{C_A}{C_F} \Bigg\{ 1 - \sqrt{ \frac{\alpha_s C_A}{18\pi}  } 
	\left(  1+ 2 \frac{n_f T_F}{C_A} - 4 \frac{n_f T_F C_F}{C_A^2}  \right)	\ + \ \mathcal{O}\left(  \alpha_s\right) \Bigg\}
\label{eq:multiplicity}	
\end{equation}
During filtering, once we re-cluster the constituents of the jet with a small radius $R_\text{filter}$, we expect to get a larger number of subjets for gluons. Since filtering does not care about the $p_T$ distribution of the subjets and simply removes all except $n_\text{filter}$ number of hardest subjets, we expect the gluon initiated jets to lose more in mass. This fact is reflected in Fig.~\ref{fig:2}, where we see a larger number of gluon-initiated jets with the degree of grooming at around $0.8-0.9$, whereas a relatively large number of quark-initiated jets keep their masses even after filtering, suggesting that $3$-hardest subjets with $R=0.3$ contain essentially all of the hard components  in quark jets.

\begin{figure}[h]
    \centering
    \includegraphics[width=\textwidth]{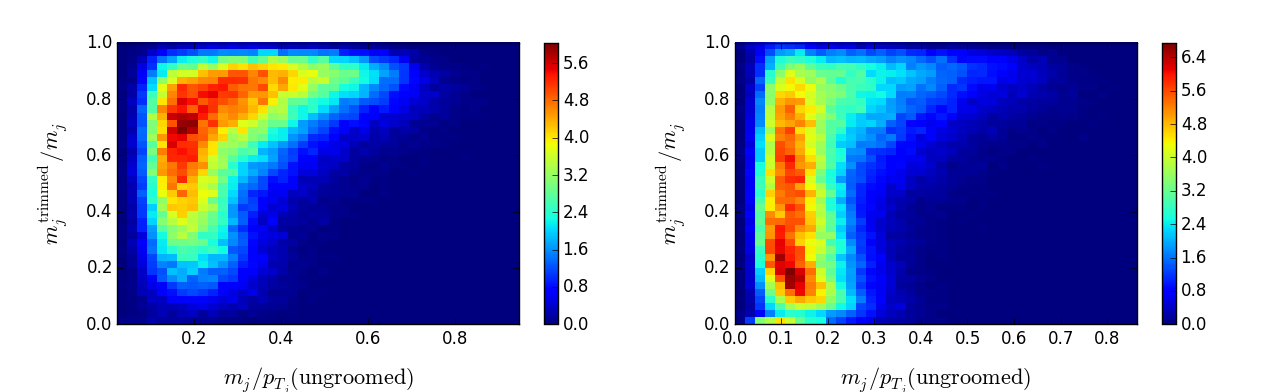}
    \caption{\label{fig:3}The effect of trimming on jet mass distributions on the gluon (left) and quark (right) initiated jets. The pdfs in each cases are plotted as functions of  $m_j/p_T$ of the ungroomed jets and the mass fraction $m^\text{trimmed}_j/m_j$.}.
\end{figure}
We obtain qualitatively and quantitatively  different effects for trimming. Because of the use of a smaller radius, namely $R_\text{trim} < R_\text{filter}$, we probe subjets of much smaller sizes (even though we are using $k_T$ algorithm instead of C/A) for trimming. Also, a subjet-$p_T$ dependent grooming procedure allows us to groom more aggressively overall compared to filtering. This explains why both $q$ and $g$ jets lose more in mass due to trimming as opposed to filtering. In order to understand the more aggressive nature of trimming in case of quark jets, note that gluon initiated  jets have a larger relative contribution from the single hard emission configuration, which is little impacted by trimming. Indeed, a pattern similar to this has also been reported by Ref.~\cite{Altheimer:2013yza}, where gluon-initiated jets are found to be less \textit{volatile}~\cite{Ellis:2012sn} under pruning~\cite{Ellis:2012sd, Ellis:2012zp}. In Fig.~\ref{fig:4} this fact is demonstrated for quark and gluon jets. A large fraction of quark jets sustain  significantly more mass loss (lose around $80\%$  of their ungroomed masses) than the gluon jets, most of which lose around $20\%$-$40\%$.  

\begin{figure}[!h]
    \centering
    \includegraphics[width=1.0\textwidth]{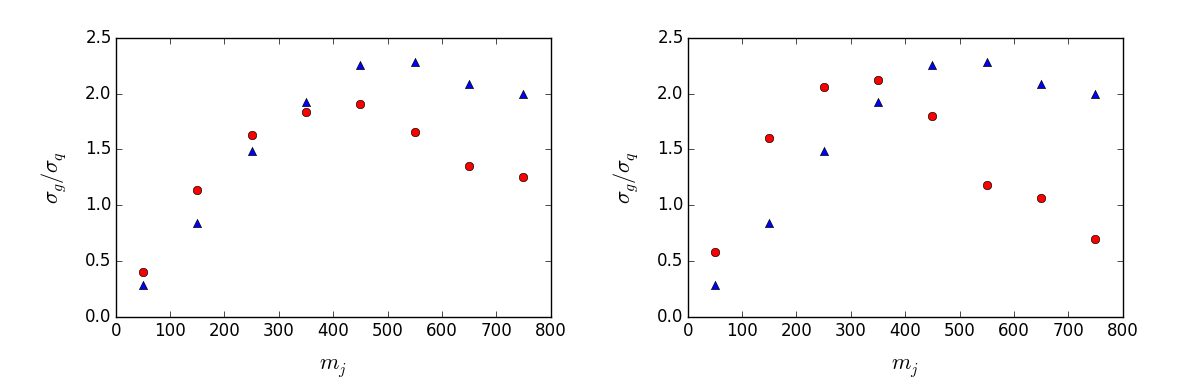}
    \caption{\label{fig:4}The $g/q$ fractions (defined in the text) as functions of jet-masses. The blue (triangle) points represent the distributions when ungroomed masses are used to populate mass bins. The distributions for 
 groomed masses are shown by  the red (circular) points. The left (right) plot uses filtering (trimming) as  the grooming algorithm.}
\end{figure}
Finally, in Fig.~\ref{fig:4} we show that grooming alters the nature of jets occupying a certain jet mass bin. We plot the $g/q$ fraction (namely, the number of gluon jets in the bin divided by the number of quark jets in the same bin) as a function of the center of the mass bin. The blue (triangle) points represent the $g/q$ fractions before jets are groomed, whereas the red (circular) points represent the same after grooming. The left (right) plot shows the result when  filtering (trimming) is used for grooming.  In case of filtering, it is straightforward to see that as the gluon initiated jets  lose more in mass, more gluons start occupying the low mass bins. As a result the gluon fraction increases for lower mass and the quark fractions increase for high mass bins.  The effect is more pronounced for trimming, suggesting that the gluon-initiated jets at high masses also lose more masses than the quarks. Note that the quark-initiated jets that lose the largest fraction of their mass via trimming originally had small ungroomed jet masses, i.e. they occupied the first few bins in the ungroomed jet mass distribution (see Fig.~\ref{fig:1} ). After trimming, these jets still occupy the first bin even if their masses have been drastically reduced. On the contrary, there are a lot more gluon jets in the high mass bin for the ungroomed case, which now move to the lower bins after grooming is done, increasing the gluon fraction in the low mass bins.  

The purpose of these plots (especially Fig.~\ref{fig:4}) is not to give a quantitive measure of the $g/q$ fraction, but rather to demonstrate that the grooming procedure introduces a bias, which is not typically accounted for in collider studies. This bias depends on specific grooming procedures, as well as the mass bins concerned. 

\begin{figure}[h]
    \centering
    \includegraphics[width=\textwidth]{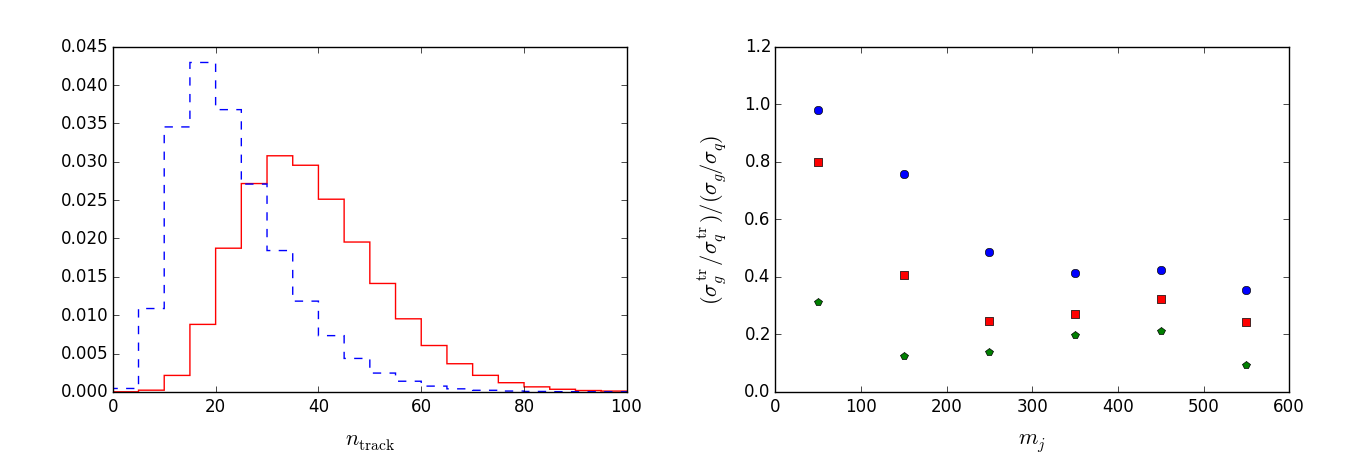}
    \caption{\label{fig:5}The effect of cuts on $n_\text{track}$, the number of tracks associated with a jet. The left figure shows the distribution of $n_\text{track}$ (each track with $p_T > 0.5\gev$) for the gluon jets (red and solid), and the quark jets (blue and dashed). The right plots show the effect of cuts on $n_\text{track}$ on the $g/q$ fraction, where we plot the double-ratio (namely, the $g/q$ fraction with a cut divided by the $g/q$ fraction without a cut). The blue (circular), red (square), and green (pentagon) points represent  $n_\text{track} < 40$, $n_\text{track} < 30$, and $n_\text{track} < 20$ respectively.}
\end{figure}
A well studied substructure variable that has been  employed in order to discriminate $q/g$ is counting the number of tracks associated with jets. The number of tracks counted inside a jet or $n_\text{track}$ is related to the number of charge particles associated with a shower and is given in Eq.~\eqref{eq:multiplicity}.	

Studying the distribution of $n_\text{track}$ is, however, troublesome. This observable is infrared unsafe, and the distributions produced by various parton-showers do not typically match~\cite{Aad:2014gea}. As we mention before, the purpose of this work is not to give a quantitive estimate of the effect of cuts on $n_\text{track}$, but rather to point out the bias introduced by a cut. In the left plot in Fig.~\ref{fig:5} we show the distribution of the number of tracks with $p_T > 0.5\gev$ associated with quarks and gluon jets (as produced by Pythia~8). The red (solid) and  blue (dashed) lines gives the distributions for the gluon and quark initiated jets respectively. In the right plot we introduce cuts on $n_\text{track}$. We measure the $g/q$ fractions in various mass-bins (ungroomed jet mass) before and after we impose the cut on $n_\text{track}$, and plot the ratios of these fractions. In the figure  we represent these double ratios  for cuts $n_\text{track} < 40$, $n_\text{track} < 30$, and $n_\text{track} < 20$ with blue (circular), red (square), and green (pentagon) points. It is straight forward to understand that  a harsh cut on $n_\text{track}$ reduces the $g/q$ fractions in each bin. A non-trivial feature is that the fractional increase in the quark content depends sensitively on the bin, suggesting that the high mass gluon jets are characterized by relatively  large numbers of tracks.

\section{\label{sec:3}Scales on the Tail of QCD Distributions due to $q/g$ Bias}
In this section we provide a concrete example where $q/g$ bias in the analysis gives a non-trivial shape to the QCD background. The di-boson resonance search by ATLAS~\cite{Aad:2015owa} provides us with the case study. As explained in the introduction, the analysis relies on the following strategy: ($i.$) collect all events with two jets; ($ii.$) employ a set of standard kinematic cuts that screen events further;  ($iii.$) subject each jet from the selected event to substructure analyses, which attempt to tag the jet to be a $W/Z$-jet (meaning, the jet includes all decay products of $W/Z$ particles); and finally ($iv.$) select all events with two tagged jets, and search for a $VV$ ($WW, WZ$, or $ZZ$) resonance in the dijet mass spectrum. 

A $VV$-resonance candidate with mass say $2\tev$ will show up in the dijet mass-spectrum as a  bump on a falling spectrum (due to QCD) at around 
$2\tev$. Naively, one expects the shape of the background to be smoothly falling. In fact, the analysis in Ref.~\cite{Aad:2015owa} relies on this. The ATLAS collaboration fits the background with a smooth function, where the slope of the jet mass distribution changes monotonically over the mass-scales of interest. Such an assumption is problematic. As we find in this study, the high mass bins in the dijet spectrum are typically dominated by $q\, \bar{q}$-events, whereas the low mass bins are mostly $g\!+\!q/\bar{q}$-events, which implies that there must be a scale where both are comparable.  Below this scale, the slopes of the falling distribution is determined by the $g\!+\!q/\bar{q}$-events, and above it the slope is given by $q\,\bar{q}$-events.  Therefore, one finds that even though the combined distribution asymptotically (far away where these two distributions cross over)  matches to individual distributions, a bump like shape may be generated where both subprocesses are comparable. 

An additional issue that can play a vital role in determining the shape of the distribution is the relative mass mis-measurement of  $q$ vs. $g$ initiated jets.  The dijet mass bin where the  $g\!+\!q/\bar{q}$ and $q\,\bar{q}$ events cross-over (hence, the location of the feature in the spectrum) depends crucially on the amount of relative mass mis-measurements. 

In this section, we begin with the details of the simulation in Subsection~\ref{subsec:3.1}, follow it up by brief descriptions of  conventional+substructure variables in Subsections~\ref{subsec:3.2} and \ref{subsec:3.3} respectively, show how we model relative mass mis-measurements for $q$ vs. $g$ initiated jets in Subsection~\ref{subsec:3.4}, and finally show the effect of these variations in the dijet mass spectrum  in Subsection~\ref{subsec:3.5}.

\subsection{\label{subsec:3.1}Simulation details}
In this subsection, we  lay out clearly the simulation details and the flow of cuts we use to come to the conclusion. 
\begin{enumerate}
\item In our study all the events are generated using Pythia 8~\cite{Sjostrand:2006za, Sjostrand:2007gs}. In order to populate a large number of QCD dijet events in the region of interest without generating an astronomical number of initial events, we impose a couple of harsh cuts at the parton level, $(i)$ on the transverse momenta of the partons; and $(ii)$ on the invariant mass of the dijet system. In particular, we impose the following criteria:
\begin{equation}
\hat{p}_T  \ > \  500\gev \qquad \text{ and } \hat{M} \ > \ 1000\gev \; . 
\end{equation}	
Both of these cuts introduce additional scales in our theory (namely, $500\gev$ for individual jet-scales and $1000\gev$ on the dijet masses), though these scales are far from the region of interest and, therefore, should not affect the analysis.

\item In order to provide a semi-realistic environment for high energy collisions, we use Delphes~\cite{deFavereau:2013fsa}. We use the standard Delphes card to simulate the details of the  ATLAS detector. We only collect the track and tower outputs from Delphes, and additional functionalities such as jet reconstruction or energy rescaling of Delphes are not used in our study. At the tower level, all entries of $p_T < 1\gev$ and associated with the hadronic calorimeter are discarded. For the electromagnetic calorimeter, we only discard tower entries of $p_T < 0.5\gev$. At the level of detector simulation, all tracks with $p_T > 0.1\gev$ are kept with varied $\eta$ and $p_T$ dependent efficiencies. For all charged particles, their respective efficiencies are maximized for $\left| \eta \right| < 1.5$ and $p_T > 1\gev$. For further details, see the Delphes card in~\cite{deFavereau:2013fsa}.

 \item The tower entries from Delphes are checked and reweighted to make sure that the $4$-vectors are massless. Further, since we are only going to restrict ourselves to the output from the central part of the detector (following Ref.~\cite{Aad:2015owa}), we only keep the tower and track outputs within $\left| \eta \right| < 2.0$. Following Ref.~\cite{Aad:2015owa}, we
impose a stronger cut on the tracks (namely, $p_T < 0.5\gev$).  After the selection is made, each track is replaced by a ``ghost" $4$-vector with arbitrarily small energy, and collinear with the corresponding track. The negligible energy of the ghost-particles ensures that, even if these are included in the clustering procedure, the jet-properties remain unaltered. Selected towers and the ghosts are clustered into jets using C/A jet algorithm~\cite{Dokshitzer:1997in, Wobisch:1998wt} as implemented in Fastjet~\cite{Cacciari:2005hq, Cacciari:2011ma}. We use $R=1.2$ and $p_{T_\text{min}} = 20\gev$ to define the jets. A straightforward counting of the number of ghosts clustered into a jet gives a count of the number of tracks associated with that jet. 

\end{enumerate}

\subsection{\label{subsec:3.2}Standard Kinematic Cuts} 

Events are selected as long as the two leading jets (namely, $J_1$ and $J_2$ with the convention $p_{T_{J_1}} \geq p_{T_{J_2}}$) from the event satisfy the following criteria:
\begin{equation}
\begin{gathered}
	p_{T_{J_1}} \ \geq \ 540\gev \qquad \text{ and }  \qquad p_{T_{J_2}} \ \geq \ 20\gev \\
	 \left| \eta_{J_1} - \eta_{J_2} \right| \ < \ 2.0 \\
	 \frac{ p_{T_{J_1}}  - p_{T_{J_2}} } { p_{T_{J_1}}  + p_{T_{J_2}} } \ < \ 0.15
\end{gathered}	 
\label{eq:basiccuts}
\end{equation}
Throughout this study, we do not alter this choice of kinematic cuts. All events that fail to meet these criteria are discarded, and all distributions presented in this paper belong to events that pass this set of cuts.   

\subsection{\label{subsec:3.3}Substructure Analysis} 
Next, both $J_1$ and $J_2$ are subject to mass-drop and filtering criteria~\cite{Aad:2015owa}. We utilize the mass-drop + filtering code as implemented in Fastjet. This algorithm, proposed originally in Ref.~\cite{Butterworth:2008iy} for finding the Higgs scalar,  de-clusters a given jet (constructed using a recombination algorithm) until it reaches a stage  of clustering where both the parents are significantly lighter than the daughter, and, at the same time, the parents have fairly similar transverse momenta. Quantitatively, this stage of mass-drop is characterized by a splitting $1+2 \rightarrow 3$, with
\begin{equation} 
\text{max}\left( m_1, m_2\right) \ \leq \ \mu_{\text{cut}} \ m_3 \qquad \text{ and } \qquad 
	 \text{min}\left( p_{T_1}^2, p_{T_2}^2 \right)  \Delta R^2_{12} \  > y_{\text{cut}} m_3^2 \; .
\end{equation}
We declare a jet to be a \textit{jet-with-substructure} if it passes the mass-drop+asymmetry criteria. All passed jets are then filtered.   During filtering, a jet's constituents are reclustered with the C/A jet algorithm with $R= R_\text{filter}$ parameter and only the $n_\text{filter}$ hardest subjets are retained. Following Ref.~\cite{Aad:2015owa} we use the following parameters
\begin{equation}
	y_\text{cut} \ = \ \left(0.45 \right)^2\, , \qquad
	\mu_\text{cut} \ = \ 1.0 \, , \qquad
	R_\text{filter} \ = \ 0.3 \, , \qquad
	n_\text{filter} \ = \ 3 \, . 
	\label{eq:mdfiltcuts}
\end{equation}
We declare a  \textit{jet-with-substructure} to be $W/Z$-tagged if the filtered mass of the jet lies in the signal window (namely, $(60-110)\gev$).  
Also note that because of the choice of a trivial $\mu_\text{cut}$ parameter, the mass drop+asymmetry cut reduces to simply an asymmetry cut. 

\subsection{\label{subsec:3.4}Implementing Relative Scaling in $q$ vs.~$g$ Initiated Jets}

In this subsection we attempt to understand the effect of a \emph{relative} mis-measurement between $q$ and $g$ initiated jets. The reasons for this study are twofold: first, while Delphes includes tower-by tower resolution functions, these may be insufficient to capture quark vs. gluon jet differences at the substructure level; second, differences between quark and gluon jets have explained excesses in the past~\cite{Aaltonen:2014mdq} and are therefore worth exploring. The analyses in this section may appear to be rather naive. However, we think that even this simplistic procedure sufficiently demonstrates  that such a mismeasurement  can be important.  The details are as follows:
\begin{enumerate}
\item One jet in the $g\!+\!q/\bar{q}$ sample is chosen at random. Given the selected jet (designated by a $4$-vector) in the direction $\hat{k}$, we rescale its momentum $4$-vector in the following way:
\begin{equation}
\left( E, P \hat{k}\right) \ \rightarrow \ \left( E(1+\delta), P(1+\delta) \hat{k}\right)   \qquad \Rightarrow \qquad \delta_m \ =  \delta \times m
\label{eq:rescale}
\end{equation}
The naive rescaling in Eq.~\eqref{eq:rescale} represents a bias in the mis-measurement of energy. It is rather simplistic in that it assumes that the energy measured in all calorimeter cells in a jet gets mis-measured by the same amount. Even though the angular information of each cell is kept unaltered, the enforcement of masslessness condition (for each cell) forces us to rescale the magnitude of momentum by the same amount. As a result, the final jet $4$-momentum is collinear to the unscaled version. A more general procedure, where each cell is rescaled independently, also changes the direction of the $3$-momentum of the jet.

\item We select $\delta$ from a normal probability distribution with mean $\langle \delta \rangle $ and standard deviation $\sigma_\delta = \langle \delta \rangle$.
\begin{equation}
P(\delta) \ = \  \frac{1}{\sqrt{2 \pi \sigma_\delta}}  
	\exp{\Big( \frac{\left( \delta - \langle \delta \rangle \right)^2}{2\sigma^2_\delta} \Big)}
\end{equation}
\item We repeat this procedure for both the jets in the $gg$-event sample. The energy and momenta for the jets in the $qq$-sample are not re-weighted. 
\end{enumerate}

\subsection{Results}
\label{subsec:3.5}

In this subsection, we describe results after an event goes through all the procedures outlined above. Before proceeding, though, let us summarize the analysis chain:  
\begin{enumerate}
\item We generate events using Pythia~$8$, and simulate the detector using Delphes. Calorimeter cells form Delphes are clustered using C/A algorithm, $R=1.2$ (see Subsection~\ref{subsec:3.1} for details). 
\item We enforce standard kinematic cuts that accept events with at least two hard jets with the leading jet $p_T > 540\gev$ and the other jet being not too dissimilar (see Eq.~\eqref{eq:basiccuts}). 
\item Each jet goes through the substructure analyses, and therefore is characterized by ($i.$) the original momentum $4$-vector,  ($ii.$) the momentum $4$-vector after grooming is done, and ($iii.$) the number of tracks associated with the ungroomed jet. Additionally, we also obtain a boolean associated with a jet (whether or not it is a \textit{jet-with-substructure}). For details see  Subsection~\ref{subsec:3.3}.  
\item Finally, jets are rescaled as outlined in Subsection~\ref{subsec:3.4}. 
\end{enumerate}

We now investigate kinematic distributions for the partonic subprocesses $p+p \to g+g$,  $p+p \to g + q/\bar{q}$, and $p+p \to q+\bar q$. We are interested in the shape and relative rates of the subprocess and study how they are affected by the substructure analyses.

\begin{figure}[h!]
    \centering
    \includegraphics[width=0.455\textwidth]{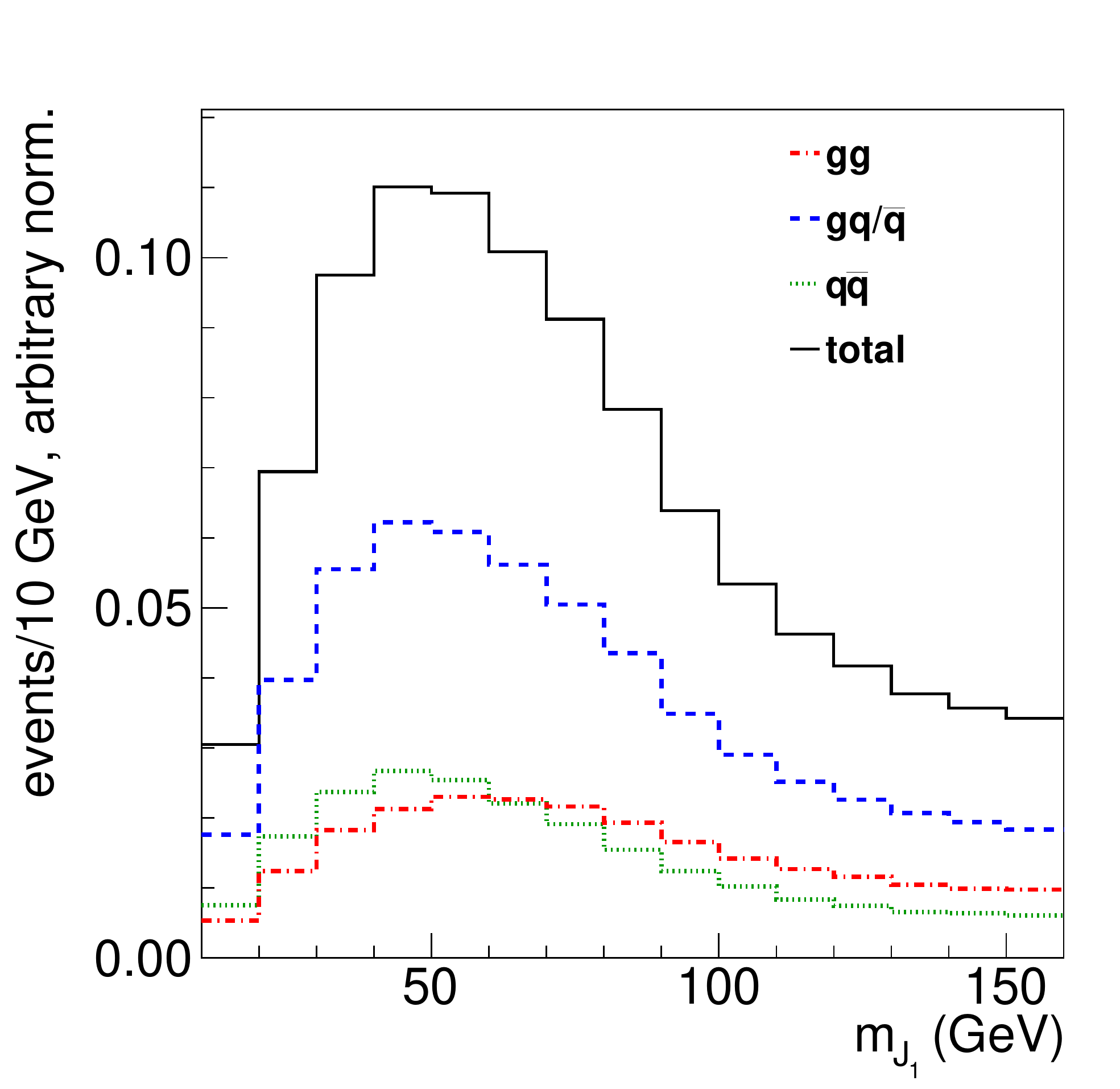}
    \includegraphics[width=0.455\textwidth]{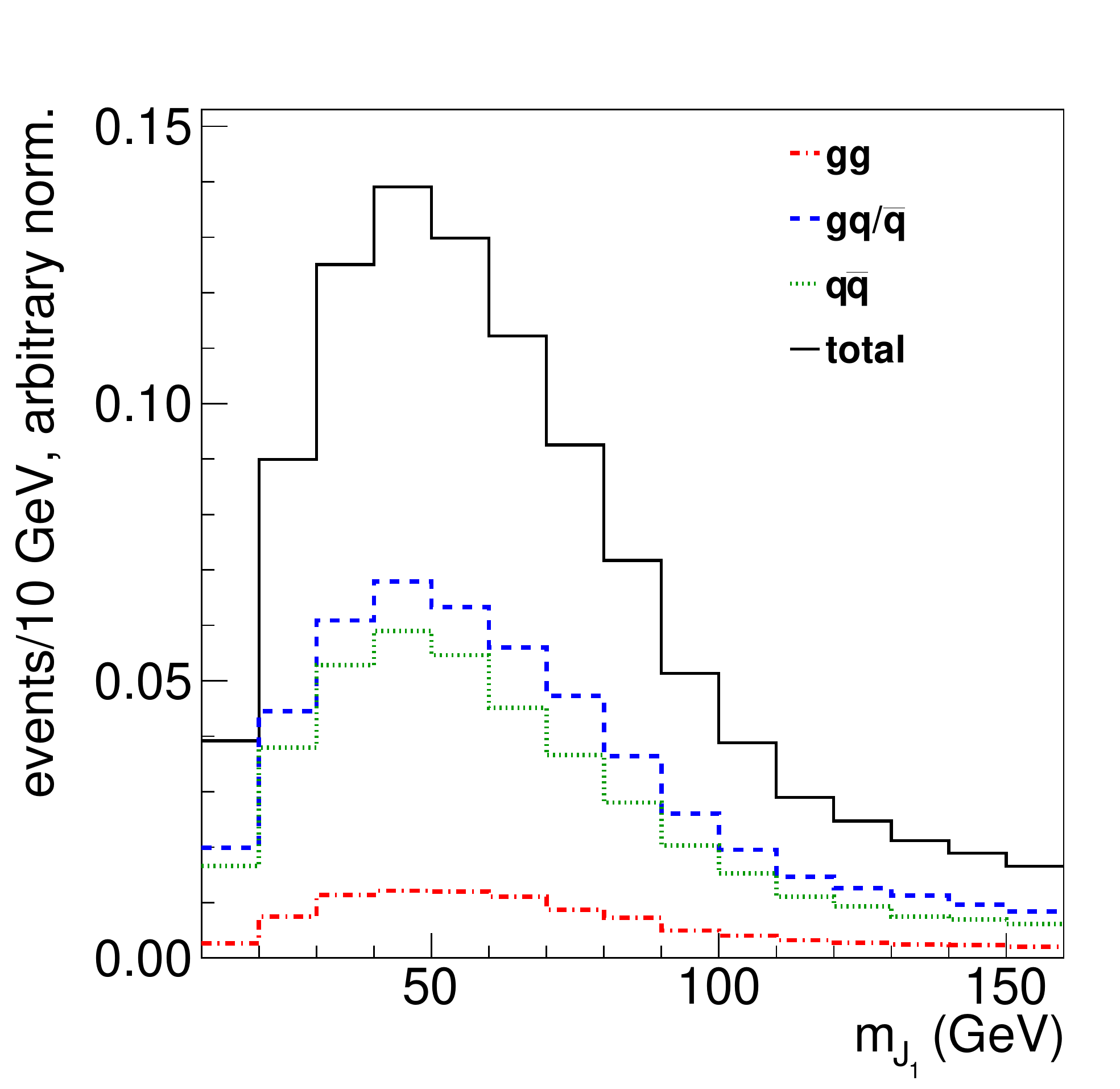}
    \caption{The mass of the leading jet ($m_{J1}$) in the dijet events, which satisfy Eq.~(\ref{eq:basiccuts}), for various partonic processes. Also, all jets in these figures are  \textit{jets-with-substructure}. The left plot contains the distribution of  ungroomed jet masses when no cut on tracks is given. The right plot shows the distributions for filtered jet masses for jets with $n_\text{track} \leq  30$.}
    \label{fig:6}
\end{figure}     
We first attempt  to understand how the number of tagged-jets (recall jets are considered tagged if their groomed mass falls within $60-110\, \gev$) depends on filtering and the $n_{{\rm track}}$ cut. We are interested in scenarios with $m_{J_1J_2} \gg m_{J_1},m_{J_2}$, where  $m_{J_1J_2}$ refers to the dijet invariant mass. Given this hierarchy, one might naively expect that $m_{J_1J_2}$  is insensitive to fluctuations in $m_{J_1},m_{J_2}$. However, as the value of $m_{J_i}$ determines whether or not a jet gets tagged, and the dijet invariant mass distribution is calculated using only tagged jets, understanding the individual jet masses is crucial.  We demonstrate the effect of filtering and $n_{track}$ cut on jet masses in Fig.~\ref{fig:6}, where we have plotted the jet-mass distribution of the leading jets for the partonic subprocesses, $p\,p \to g\,g$,  $p\,p \to g + q/\bar{q}$, and $p\,p \to q\,\bar q$.  Rather than focusing on the signal window, we take a larger range in jet-masses. All events plotted in this figure satisfy Eq.~(\ref{eq:basiccuts}), and all jets in these figures are  \textit{jets-with-substructure}. In the left frame of this plot we show the ungroomed jet masses for the leading jets before we apply filtering or a cut on the number of tracks. In the right frame we show the distributions of filtered jet masses after we impose the cut on $n_\text{track} \leq 30$.

Focusing on the signal window of Fig.~\ref{fig:6}, we find that the nature of jets in the signal window changes drastically as we apply the cut on $n_\text{track} $ and filter the jet. In the left plot the signal window is dominated by jets from $g \!+\! q/\bar{q}$-events, whereas in the right plot a much higher fraction of jets within the signal window arises from the $q\,\bar{q}$-event sample.  Importantly, we  find that the effect of filtering on \textit{jets-with-substructure} is relatively minor. The drastic change in the nature of jets in the signal window is mostly due to cuts in asymmetry+$n_\text{track}$. This conclusion is in line with the effects shown in Fig.~\ref{fig:5}.

\begin{figure}[h!]
    \centering
    \includegraphics[width=0.455\textwidth]{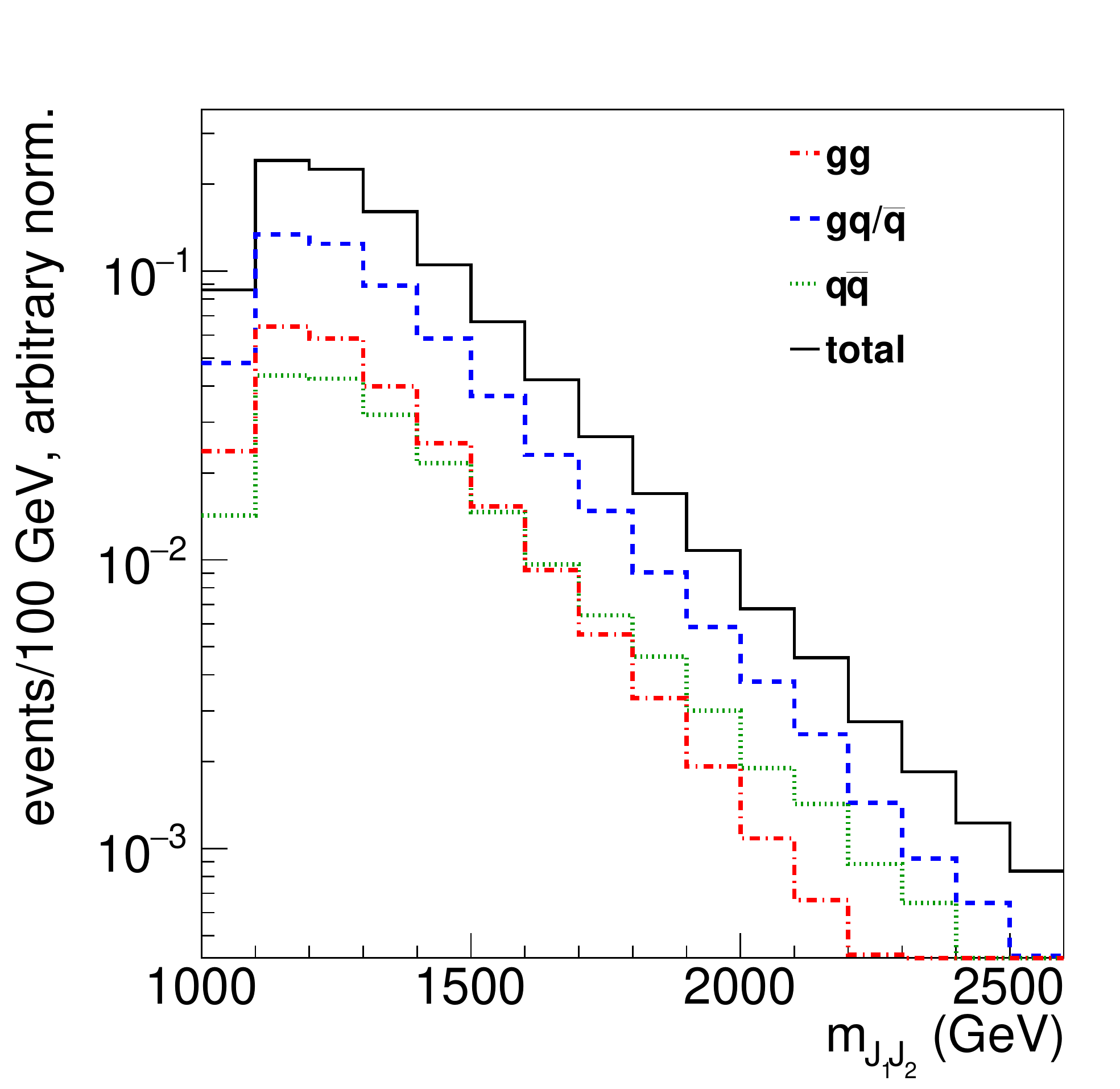}
    \includegraphics[width=0.455\textwidth]{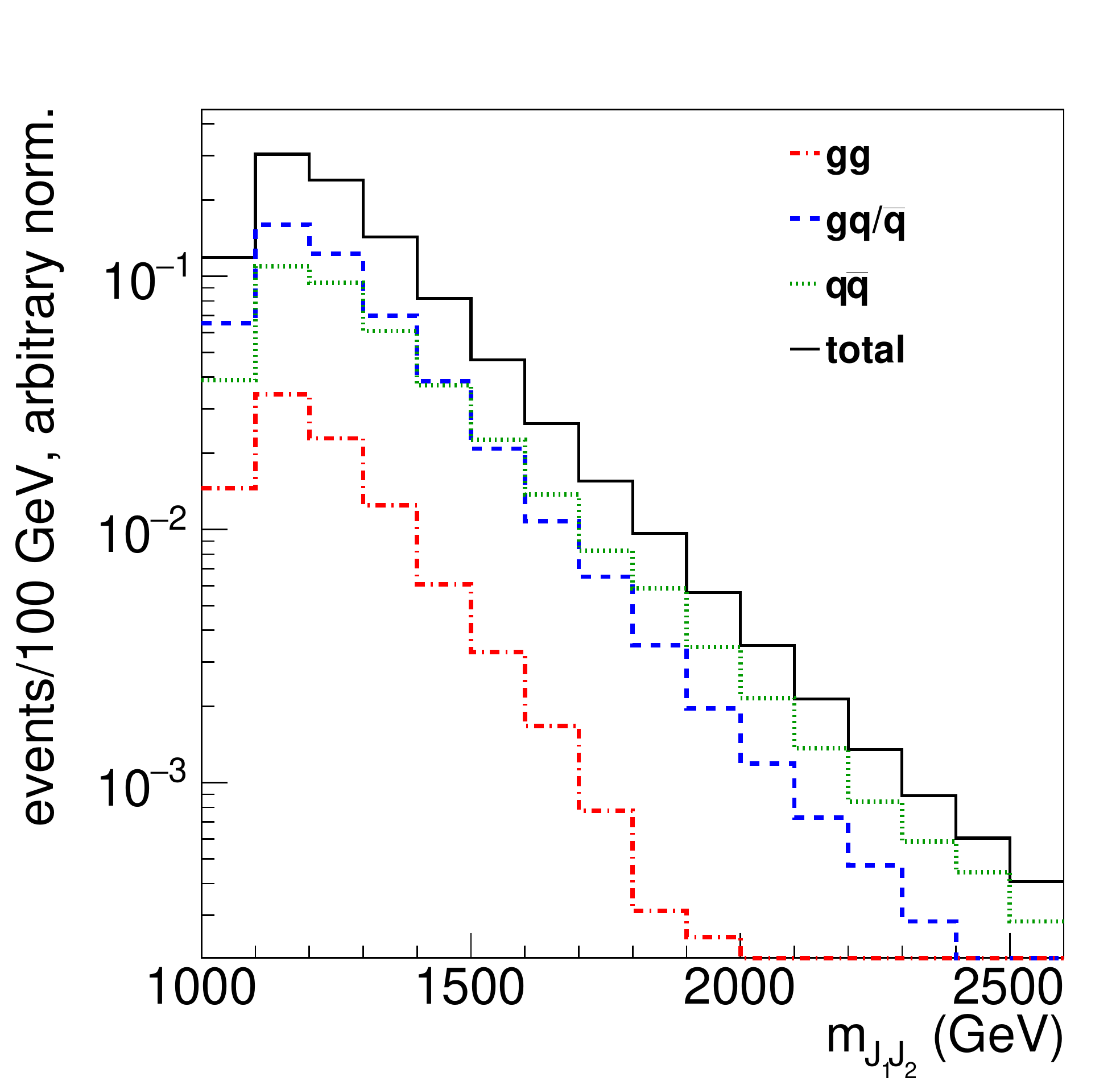}
    \caption{\label{fig:7} The $m_{J_1J_2}$ distributions for the dijets, with the different colors indicating different partonic contributions; red for $g\,g$, green for $q\,\bar{q}$, blue for $g \,+\, q/ \bar{q}$, and the total of all contributions in black. The jets in all four panels fall in the signal mass window $60\, \gev < m_J < 110\, \gev$ and pass the $p_T$, mass-drop, and jet balance cuts from Eq.~\eqref{eq:basiccuts}. The difference between panels lies in the track and filtering requirements; the left panel shows the distributions when ungroomed jets without the track requirement are used to construct the dijet invariant mass, while the right panel includes a track requirement and uses filtered jets.}
\end{figure}    
Our results in Fig.~\ref{fig:6} imply that the combination of jet grooming and a $n_\text{track}$ cut might impart a nontrivial shape on the $m_{J_1J_2}$ distribution. As this procedure has a higher acceptance for the $q\,\bar{q}$-event sample, the number of events that pass all selection criteria will have a significantly higher fraction of $q\,\bar{q}$-events. If the differential distributions in $q\,\bar{q}$-events differ significantly from that in $ g\,+\, q/ \bar{q}$-events, one may expect a new scale to arise where these partonic processes contribute equally.  
We illustrate this point in  Fig.~\ref{fig:7}, where  we show the distribution of the dijet invariant mass $m_{J_1J_2}$  for the mass range $1.0\tev \leq m_{J_1J_2} \leq 2.5\tev$.  For this plot we require that the event pass the selection criteria in Eq.~\eqref{eq:basiccuts}, that both jets in each event are \textit{jets-with-substructure}, and that both jets are tagged. In the left panel we use ungroomed jets to construct the dijet mass spectrum, whereas we only use filtered jets with $n_\text{track} \leq  30$ for the right panel.  As shown in these plots,  at low $m_{J_1J_2}$, the dijets primarily come from processes with one quark/antiquark and one gluon at parton level. However at higher $m_{J_1J_2}$,  $p\,p \to q\,\bar q$ takes over.  More importantly, the crossing point where the contributions from $q\,\bar q $ and $g + q/\bar q$ are equal is strongly sensitive to the substructure analysis and the track requirement.  We get a glimpse of this sensitivity by comparing the left and right panels of Fig.~\ref{fig:7}.  If we do not impose a cut on $n_\text{track}$, we do not find  the crossing point within the range of study. However, after requiring $n_\text{track} \leq 30$ in each jet (\emph{before} filtering), the crossing point for the filtered dijets shows up at around  $1.5\tev$ and the $g\,g$ contribution becomes negligible throughout the whole mass range. 

Figures~\ref{fig:6} and \ref{fig:7} demonstrate that the differential distribution $d\sigma_{JJ}/dm_{J_1J_2}$  for the QCD background may not be adequately described by a smoothly falling function with a slope decreasing monotonically, such as the function used by the ATLAS collaboration,
\begin{equation}
f(x) \ = \ p_0 \: x^{p_1} \left( 1 - x \right)^{p_2} \, , 
\label{eq:fit}
\end{equation}
where $p_i$ are coefficients and $x=m_{J_1J_2}/(8000\gev)$. It is more appropriate to fit the differential distributions for each of the subprocesses  with the function in Eq.~\eqref{eq:fit}. Fitting the total distribution with Eq.~(\ref{eq:fit}) makes sense only if one of the partonic subprocesses completely dominates over the entire domain\footnote{The ATLAS fit was validated in several control samples, such as before $W/Z$ tagging or using tagged jets with a signal window $40 \le m_{J_i} \le 60\, \gev$. However, unlike the signal region, the control samples are dominated by a single partonic process: before boson tagging, $g + q/\bar q$ completely dominates while $q\bar q$ dominates the $40 \le m_{J_i} \le 60\, \gev$ sideband  for $m_{J_1,J_2} > 1.2\, \tev$. } . If the domain we want to fit includes a crossing point where the dominant subprocess changes, such a simple fit will not suffice. To be more exact, we define the point of crossing (namely $\mu_\text{cross}$) via the relation:  
\begin{equation}
\mu_\text{cross} \  \rightarrow \   \Bigg( \frac{d\sigma_{qg}}{dm_{J_1J_2}}  \ -  
		\frac{d\sigma_{qq}}{dm_{J_1J_2}}  \Bigg)_{m_{J_1J_2} = \mu_\text{cross} } \ = \ 0 \; .  	
\label{eq:crosseq}
\end{equation}
For $m_{J_1J_2} \ll \mu_\text{cross}$ and for $m_{J_1J_2} \gg \mu_\text{cross}$, a function like Eq.~\eqref{eq:fit} provides an appropriate fit, with  different parameters in these two ranges.  If we insist on fitting the total background with a single function of the for in Eq.~(\ref{eq:fit}), one generically obtains a  good fit for the $m_{J_1J_2} \ll \mu_\text{cross}$ region, as this is where most of the data lies, and a mis-measured tail. With low statistics at $m_{J_1J_2} \gg \mu_\text{cross}$, one may erroneously mistake the mismodeled tail for  a bump due to new physics.

To get a more concrete idea of what features this transition scale can introduce, we turn to pseudo-experiments. Starting with a sample of dijet events which passes the kinematic and substructure cuts, we select events at random and apply a track cut and $q$ vs. $g$ smearing following Sec.~\ref{subsec:3.4}. If both jets in the selected events, post smearing and track cut, have mass in the signal window $60\,\gev \le m_{J_i} \le 110\,\gev$, we record the dijet mass. We repeat this procedure until we find 604 events, the number of events ATLAS reports using the $WW$ selection, then plot and fit (using Eq.~\eqref{eq:fit}) the distribution\footnote{We note that our signal mass window is slightly different than the ATLAS $W$ or $Z$ selection. Additionally, as we have generated events with a parton-level dijet mass cut of $1\tev$, our dijet mass distribution does not match with ATLAS at the lowest $m_{J_{1}J_{2}}$ values. To account for this, our $m_{J_1 J_2}$ fit is restricted to the range $1.3-2.6\,\tev$ }. Since the number of events is limited, the fit is driven by the low-$m_{J_1J_2}$ region where $g + q/\bar q$ dominates. This procedure has to be repeated from scratch for every choice of track cut and smearing in order to maintain the interplay between the individual jet mass window and the dijet mass spectrum. Four sample pseudo-experiments, generated using a track cut of $\le 35$ tracks and smearing with $\langle \delta \rangle = -0.03$, is shown below in Fig.~\ref{fig:bumpy}.
\begin{figure}[h!]
\centering
\includegraphics[width=0.455\textwidth]{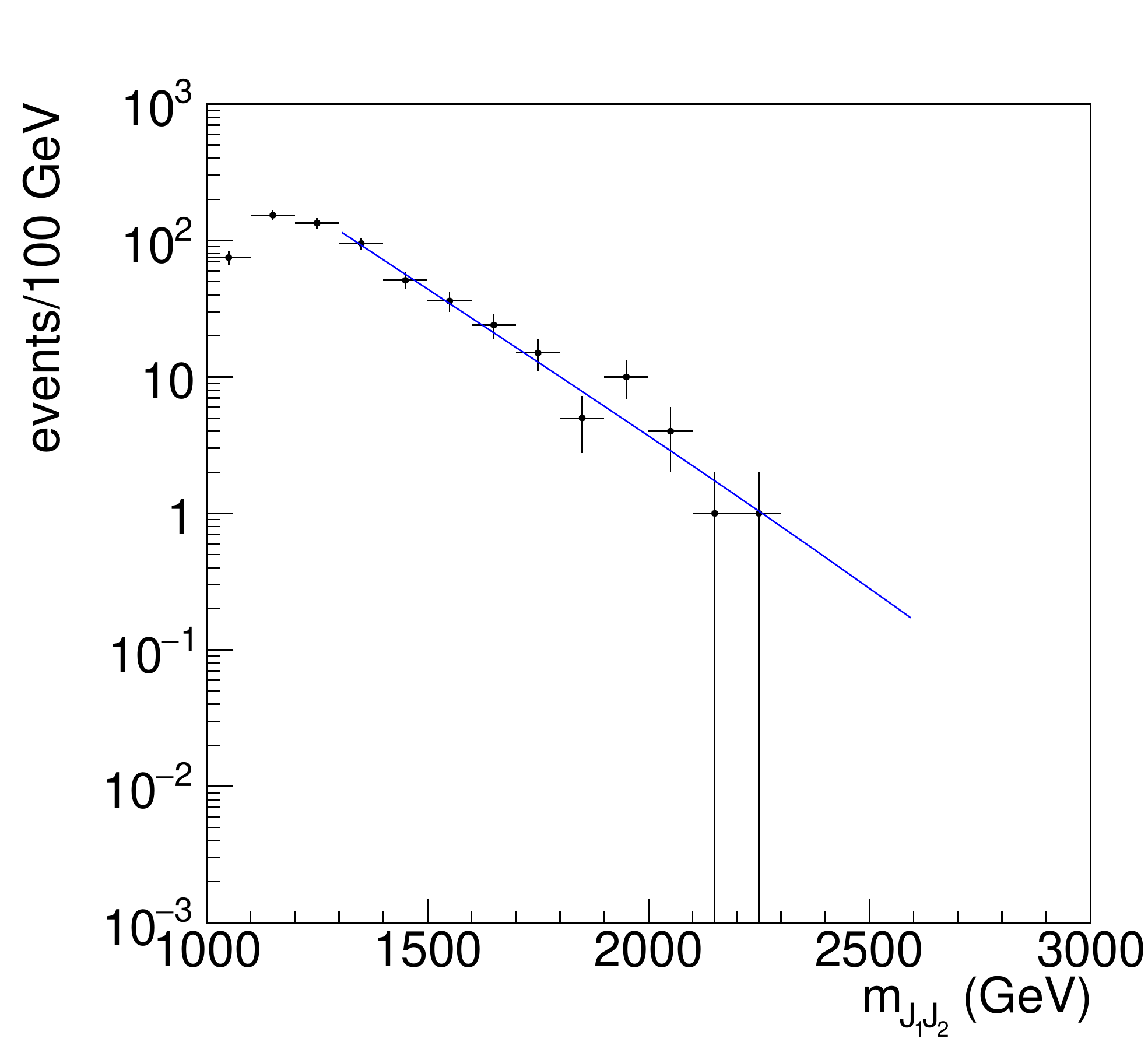}
\includegraphics[width=0.455\textwidth]{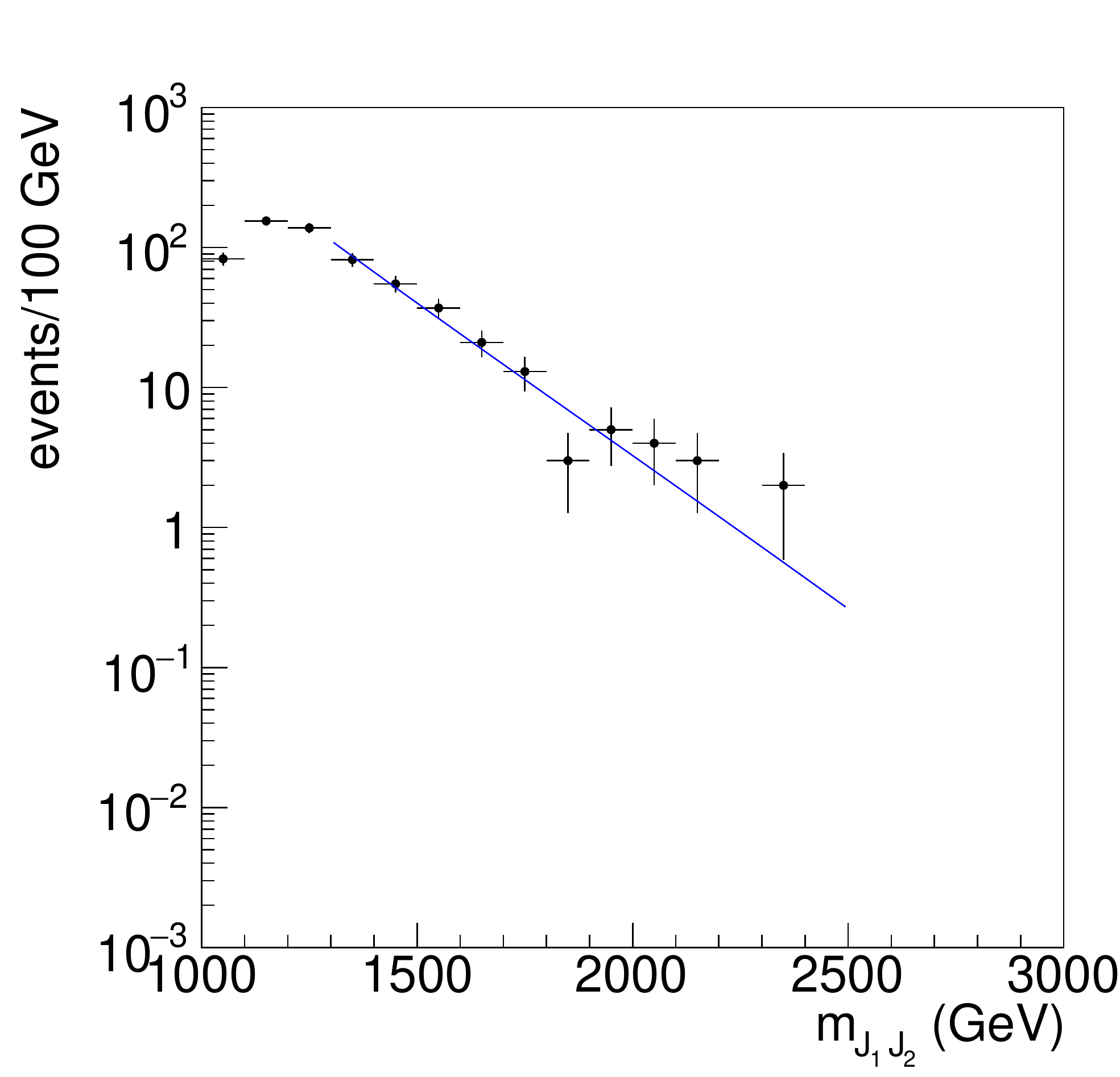} \\
\includegraphics[width=0.455\textwidth]{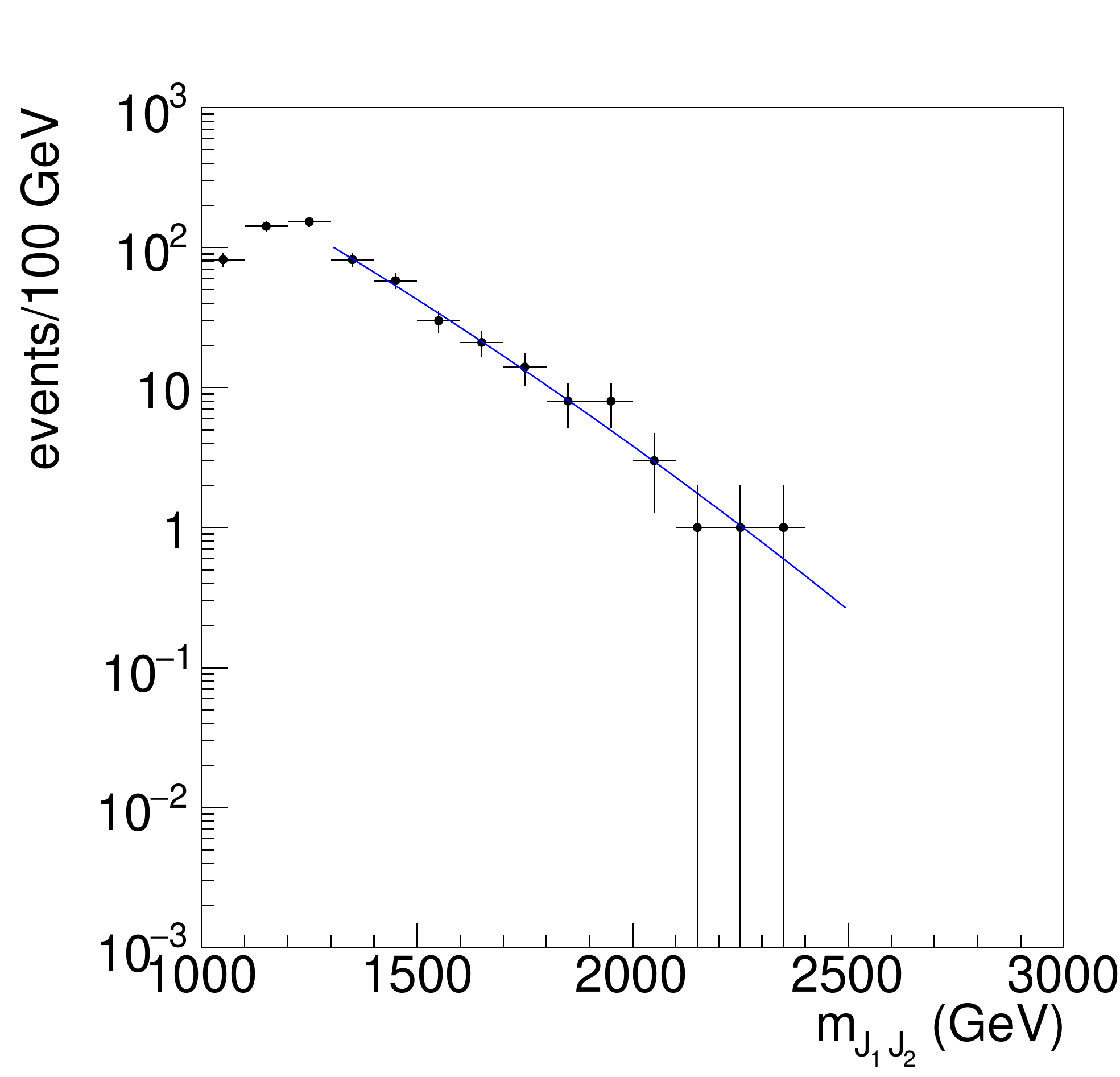}
\includegraphics[width=0.455\textwidth]{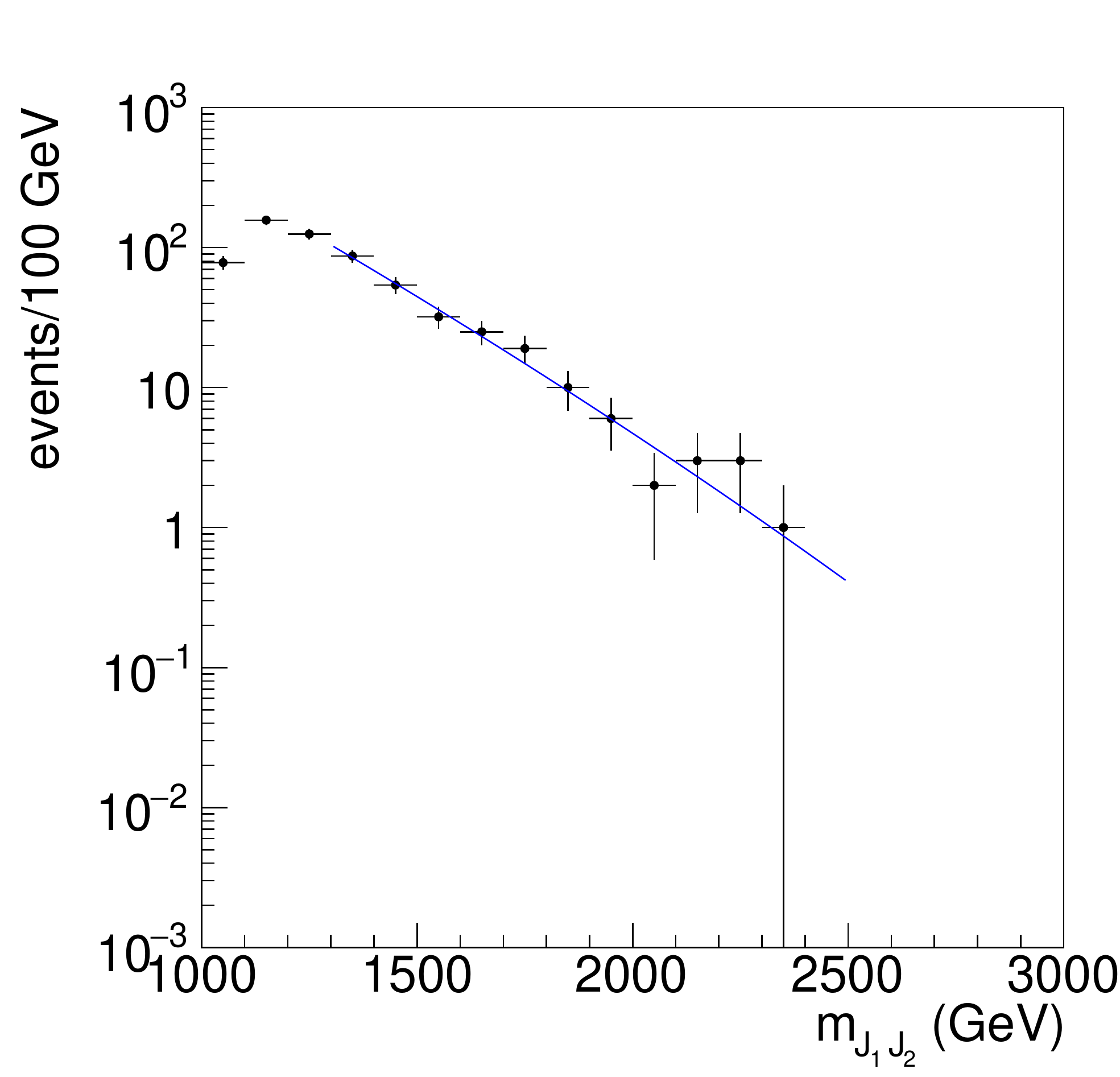}
\caption{Dijet invariant mass spectrum for four sample pseudo-experiment using a track cut of $\leq 35$ tracks, rather than the ATLAS value of $\le 30$, and smearing gluon jets with respect to quark jets with $\langle \delta \rangle = -0.03$. For this set of parameters we find $\mu_\text{cross} \sim 1.8\, \tev$, and a noticeable bump.  }
\label{fig:bumpy}
\end{figure}

While Fig.~\ref{fig:bumpy} shows just a handful of pseudo-experiments for a fixed $n_{\rm track}$ and $\langle \delta \rangle$, it does illustrate that a $\mu_\text{cross} \sim1.8\tev$ can generate the similar features in the dijet mass spectrum to what ATLAS observes in Ref.~\cite{Aad:2015owa}. Scenarios with higher or lower crossing points are less likely to look like ATLAS data. If the crossing point is higher, any features that show up as a result of $\mu_\text{cross}$ will be at $m_{J_1J_2} > 2\tev$, while if the crossing point is lower the fit would be driven by $q\,\bar q$ subprocesses and any shape differences between $g \!+\! q/\bar q$ and $q\,\bar q$ will be fitted away. The $n_{\rm track}$ used in Fig.~\ref{fig:bumpy} is different than the cut used in the ATLAS analysis, but the difference is comparable to the variation in $n_{\rm track}$ among different Monte Carlo generators~\cite{Gallicchio:2012ez, Aad:2014gea}.

In this particular scenario, therefore, understanding $\mu_\text{cross}$ is critical. However, as we have pointed out repeatedly, $\mu_\text{cross}$  depends sensitively on the substructure analysis, the cut on $n_\text{track}$, and on the  relative mis-measurement in $q/g$-initiated jets. Note that, in Figs.~\ref{fig:6} and \ref{fig:7} we do not use any relative rescaling, and only compare the variation of $\mu_\text{cross}$ after the substructure cuts have been applied and, therefore, find little dependence on filtering.  In general, however, all the three quantities play important roles. In the rest of this section, we show  the variation of $\mu_\text{cross}$ as we change these.  

Given a grooming procedure, a cut on $n_\text{track}$, and a given $\langle \delta \rangle$, we determine  $\mu_\text{cross}$ by fitting  the differential  distributions for each of the parton level processes independently with the function in Eq.~\eqref{eq:fit}.  The value of  $\mu_\text{cross}$ is calculated numerically, as the crossing point of these fitted functions.  We show the results of this study in Fig.~\ref{fig:8}, where we plot $\mu_\text{cross}$ as functions of $\langle \delta \rangle$ for different values in $n_\text{track}$.  The left and right panels of Fig.~\ref{fig:8} show the comparison of this variation as we change the center of mass c.o.m. energy of the $pp$ collision. 

\begin{figure}[h!]
\centering
\includegraphics[width=0.455\textwidth]{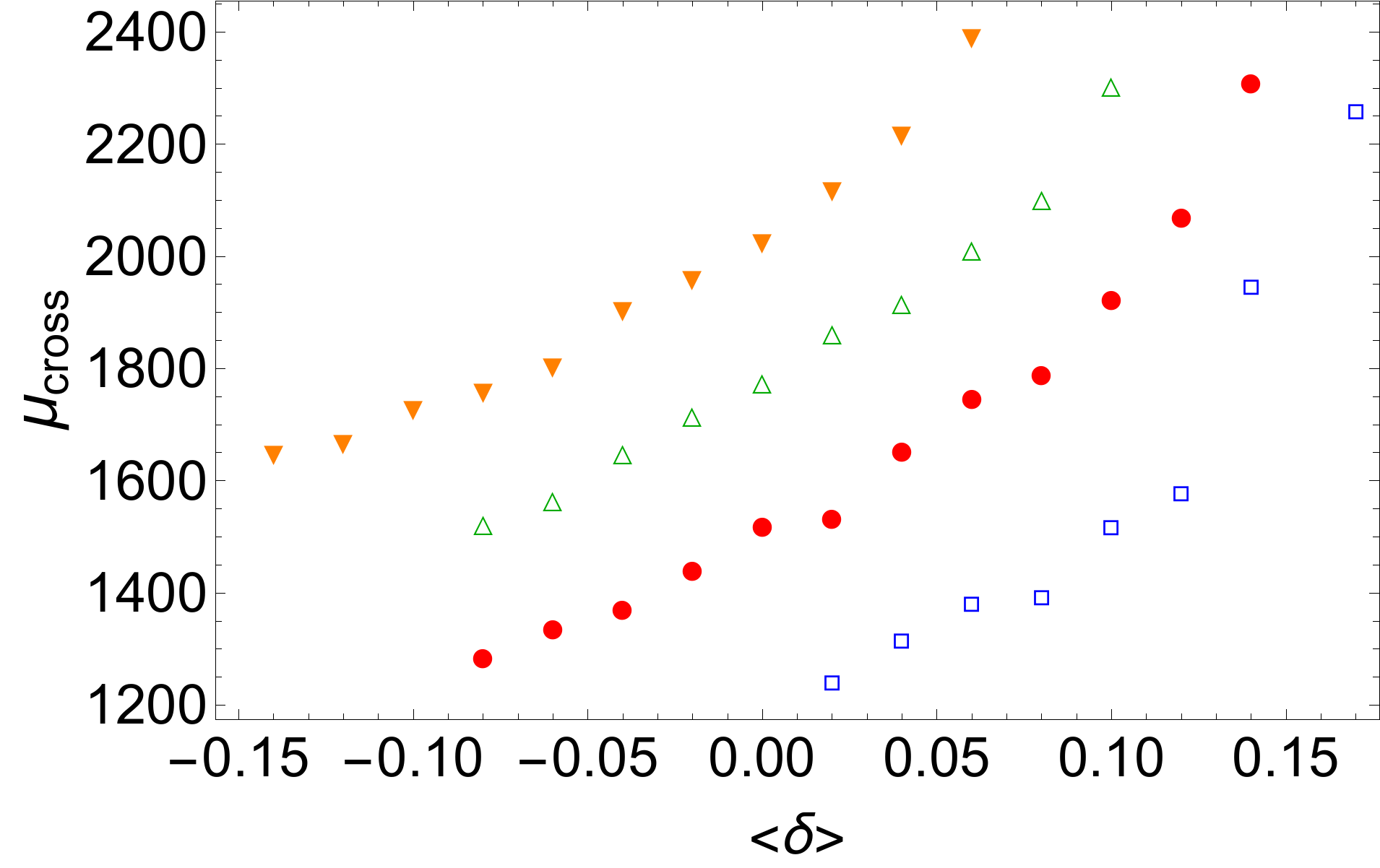} \qquad
\includegraphics[width=0.455\textwidth]{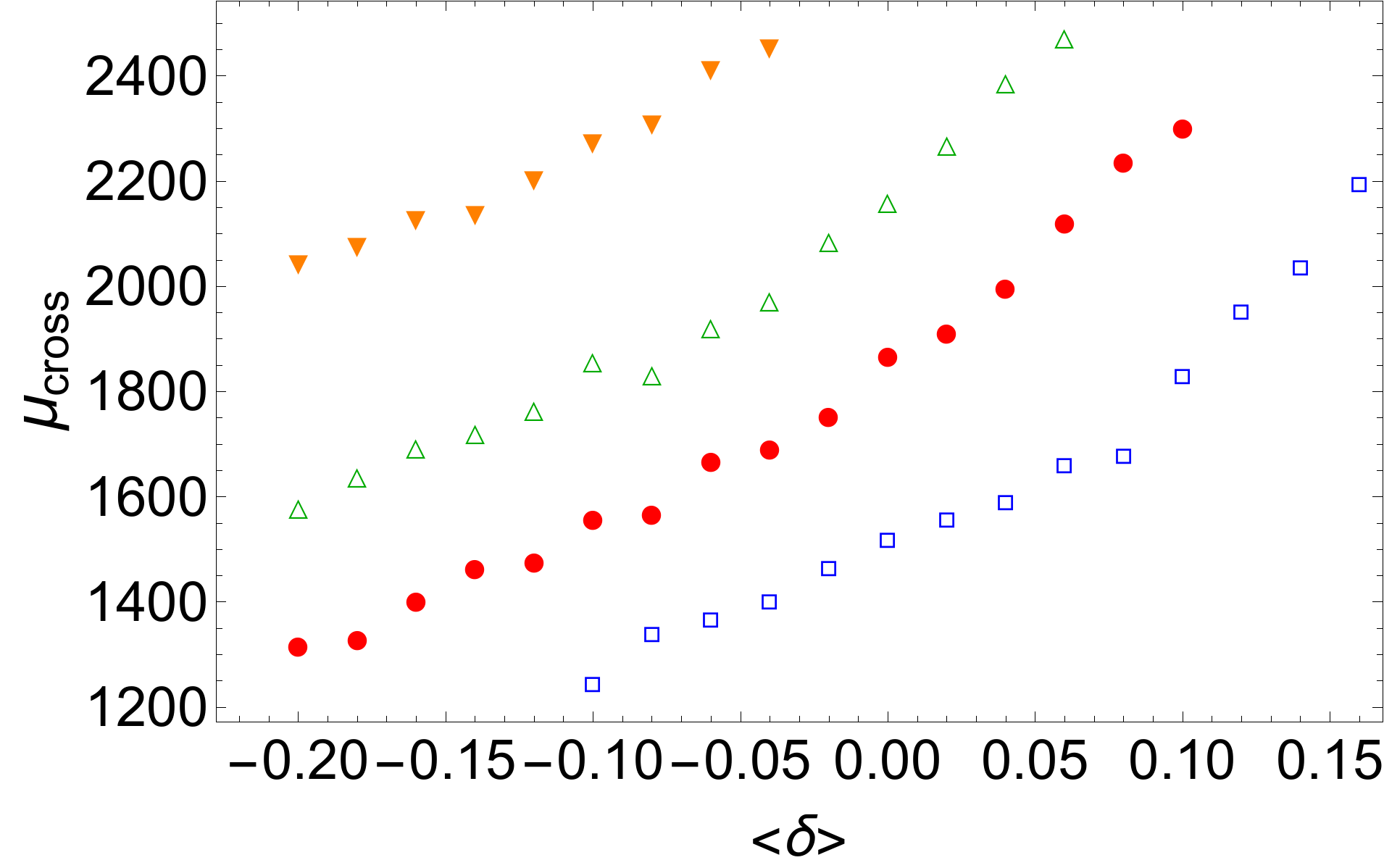}
\caption{Crossing point $\mu_\text{cross}$, defined in Eq.~\eqref{eq:crosseq} as a function of the $q/g$ energy mis-measurement parameter $\langle \delta \rangle$ for various assumptions on the number of tracks allowed. The left (right) plot shows the variation for a $8\tev$ ($13\tev$) collider. In both the plots, the blue empty squares correspond to each jet having $\leq 25$ tracks, red circles are $\le 30$, green triangles $\le 35$, and orange inverted triangles $\le 40$. The requirement in the ATLAS analysis is $\le 30$ tracks.}
\label{fig:8}
\end{figure}
The generic pattern of  the variation of $\mu_\text{cross}$  as functions of $\langle \delta \rangle$, is same whether the c.o.m. energy is $8\tev$ (left plot) or $13\tev$ (right plot). As more tracks are admitted into jets, the gluon (either in $g + q/\bar q$ or $gg$ parton processes) contribution to dijet events grows. In the absence of any $q/g$ relative mis-measurement, this pushes the crossing point out to higher values of $m_{J_1J_2}$. Smearing with a negative $\delta$ combats this trend, as negative $\delta$ implies that the gluon jet energies decrease relative to quark jets, driving $m_{J_1J_2}$ to lower values for subprocesses with gluons. By the same logic, tightening the track requirement (recall $\leq 30$ is the number used by ATLAS), suppresses the gluon contributions and brings $\mu_\text{cross}$ lower, which can be compensated by having a larger $q/g$ relative mis-measurement with positive $\delta$. For the same cuts on $n_\text{track}$, and $\langle \delta \rangle$, the crossing point $\mu_\text{cross}$  increases with higher c.o.m. energy. 

Before ending this section let us restate that, once we impose the mass drop + asymmetry cut in Eq.~\eqref{eq:mdfiltcuts}, filtering has a mild effect on the value of $\mu_\text{cross}$ as long we use the same cut on $n_\text{track}$. However, there is no reason to suspect that all grooming methods will have such a minor effect. In fact, filtering is the least aggressive groomer to begin with. One also expects a sizeable effect if we change the  substructure analysis itself (i.e. an algorithm other than mass-drop + asymmetry). In this context, we study the behavior of $\mu_\text{cross}$ if we replace the mass drop + asymmetry cut + filtering part of the analysis with trimming. Specifically, we select all events that satisfy the cuts in Eq.~\eqref{eq:basiccuts}. All jets from the selected events are then trimmed using the parameters in Eq.~\eqref{eq:groomparam} and, as before, we declare a jet to be $W/Z$-tagged if the trimmed mass of the jet lies in the signal window (namely, $(60-110)\gev$). 

Using this trimmed version of the analysis, we then study the behavior of $\mu_\text{cross}$ as we vary the number of tracks and the relative quark vs. gluon smearing. The results are shown below in Fig.~\ref{fig:9}.
\begin{figure}[h!]
\centering
\includegraphics[width=0.65\textwidth]{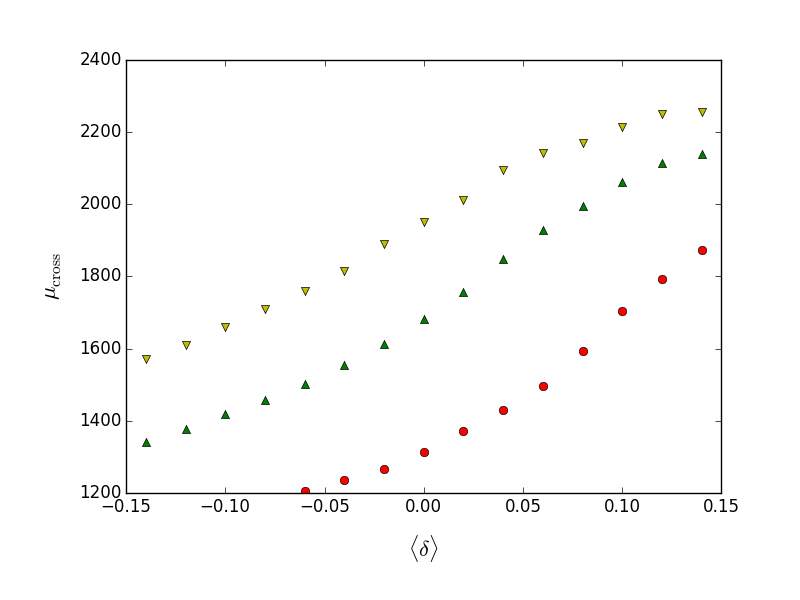} 
\caption{Crossing point $\mu_\text{cross}$ as a function of $\langle \delta \rangle$ for various assumptions on the number of tracks allowed. In the plot we use trimming as substructure variable instead of mass-drop+filtering. The  red circles correspond to each jet having $\leq 30$ tracks, green triangles $\leq 35$, and yellow inverted triangles $\leq 40$. }
\label{fig:9}
\end{figure}
The variation of $\mu_\text{cross}$ as a function of $\langle \delta \rangle$ for a given cut on $n_\text{track}$, shows a similar pattern as filtering.  However, the 
exact value of  $\mu_\text{cross}$, when all other cuts remain the same,  depends on whether we use trimming or   mass drop + asymmetry cut + filtering. For example, when we set $\langle \delta \rangle = 0.04$, $n_\text{track} \leq 35$, we find $\mu_\text{cross}$ gets lowered  from $1.9\tev$ to $1.8\tev$ as we use trimming  for substructure analysis.    

\section{\label{sec:4}Conclusion}

In resonance searches, one looks for an invariant mass bump on top of a smoothly falling background that is usually modeled by a simple monotonic function. In this paper, we question the applicability of this modeling approach in analyses that use jet substructure techniques. Using Monte Carlo, we investigate commonly utilized techniques like filtering, trimming, and a cut on the number of tracks within a jet (the last of which is infared unsafe), and find that they exacerbate the differences between quark-initiated and gluon-initiated jets. When these $q/g$ differences are propagated to more complicated observables such as the dijet mass, the result is a relative shift in which partonic subprocess ($q\,\bar q, g\,q,$ or $g \!+\! q/\bar q$ for the case of a dijet study) dominates. As different subprocesses have different shapes, the transition from one dominant subprocess to another cannot reliably be modeled by a simple monotonic function. Additionally, data-driven validation of fitting functions can be misleading if the subprocess composition in sidebands/control regions is significantly different than in the signal region.

To  assess the impact of jet substructure variables on $q$ vs. $g$ jets quantitatively, we investigate the recent ATLAS search for resonances decaying to a pair of hadronic $W/Z$~\cite{Aad:2015owa}. This analysis received a lot of attention as it revealed a tantalizing hint of an excess around $1.8-2.0\tev$. In order to better separate signal from background, the jets in the ATLAS analysis are checked to make sure that these do not contain more that a certain number of charged tracks inside, are then checked for substructure, and finally are filtered. 

Repeating the ATLAS analysis on Monte Carlo dijet events, we find that the substructure cuts along with a cut on the track-count, in particular, induce a transition scale $\mu_\text{cross}$ in the dijet mass distribution.  At masses below $\mu_\text{cross}$, $g\! +\! q/\bar q$ partonic processes dominate, while above it $q\,\bar q$ is largest. If we fit the background by a single, simple function with monotonically decreasing slope, the fit is dominated by low $m_{J_1J_2}$, where the bulk of the events lie. Extrapolating this  ($g\!+ \!q/\bar q$ driven) fit into the region where $q\,\bar q$ dominates,  the change in background slope can appear -- when viewed with limited statistics -- as an excess. Apart from depending on the cuts on substructure variables, we also find the value of $\mu_\text{cross}$ to be sensitive to any relative quark jet vs. gluon jet energy mismeasurement. Exploring these dependencies, we find there are several combinations of $n_{\rm track}$ and the relative $q$ vs. $g$ smearing parameter $\langle \delta \rangle$ that yield a crossing scale in the vicinity of $1.8\, \tev$. This value is significant because, for $\mu_\text{cross} \sim 1.8\, \tev$ we find the mismatch between the $q\,\bar q$ background and $g \!+\! q/\bar q$ fit can reproduce the excess observed by ATLAS (once the limited number of background events have been taken into account). While it still may be the case that the excess seen by ATLAS is due to new physics and will persist with more data, the results of our study demonstrate the dangers in overly simple background modeling in analyses employing jet substructure.
 
\section*{Acknowledgements}
The authors would like to thank Stephen D.~Ellis for his helpful comments during the course of this work, and while preparing the draft. This work was supported in part by the US National Science Foundation under Grant PHY- 1417118. AM thanks Boston University for computing resources.  A significant part of the computations were performed on the Mapache cluster in the HPC facility at LANL. 


\bibliographystyle{JHEP}
\bibliography{vvresonance}

\providecommand{\href}[2]{#2}\begingroup\raggedright\begin{thebibliography}{10}

\bibitem{Seymour:1993mx}
M.~H. Seymour, {\it {Searches for new particles using cone and cluster jet
  algorithms: A Comparative study}},  {\em Z.Phys.} {\bf C62} (1994) 127--138.

\bibitem{Butterworth:2007ke}
J.~Butterworth, J.~R. Ellis, and A.~Raklev, {\it {Reconstructing sparticle mass
  spectra using hadronic decays}},  {\em JHEP} {\bf 0705} (2007) 033,
  [\href{http://xxx.lanl.gov/abs/hep-ph/0702150}{{\tt hep-ph/0702150}}].

\bibitem{Brooijmans:1077731}
G.~Brooijmans, {\it High pt hadronic top quark identification},  Tech. Rep.
  ATL-PHYS-CONF-2008-008. ATL-COM-PHYS-2008-001, CERN, Geneva, Jan, 2008.

\bibitem{Butterworth:2008iy}
J.~M. Butterworth, A.~R. Davison, M.~Rubin, and G.~P. Salam, {\it {Jet
  substructure as a new Higgs search channel at the LHC}},  {\em
  Phys.Rev.Lett.} {\bf 100} (2008) 242001,
  [\href{http://xxx.lanl.gov/abs/0802.2470}{{\tt arXiv:0802.2470}}].

\bibitem{Butterworth:2009qa}
J.~M. Butterworth, J.~R. Ellis, A.~R. Raklev, and G.~P. Salam, {\it
  {Discovering baryon-number violating neutralino decays at the LHC}},  {\em
  Phys.Rev.Lett.} {\bf 103} (2009) 241803,
  [\href{http://xxx.lanl.gov/abs/0906.0728}{{\tt arXiv:0906.0728}}].

\bibitem{Kaplan:2008ie}
D.~E. Kaplan, K.~Rehermann, M.~D. Schwartz, and B.~Tweedie, {\it {Top Tagging:
  A Method for Identifying Boosted Hadronically Decaying Top Quarks}},  {\em
  Phys.Rev.Lett.} {\bf 101} (2008) 142001,
  [\href{http://xxx.lanl.gov/abs/0806.0848}{{\tt arXiv:0806.0848}}].

\bibitem{Thaler:2008ju}
J.~Thaler and L.-T. Wang, {\it {Strategies to Identify Boosted Tops}},  {\em
  JHEP} {\bf 0807} (2008) 092, [\href{http://xxx.lanl.gov/abs/0806.0023}{{\tt
  arXiv:0806.0023}}].

\bibitem{Almeida:2008yp}
L.~G. Almeida, S.~J. Lee, G.~Perez, G.~F. Sterman, I.~Sung, et~al., {\it
  {Substructure of high-$p_T$ Jets at the LHC}},  {\em Phys.Rev.} {\bf D79}
  (2009) 074017, [\href{http://xxx.lanl.gov/abs/0807.0234}{{\tt
  arXiv:0807.0234}}].

\bibitem{Plehn:2009rk}
T.~Plehn, G.~P. Salam, and M.~Spannowsky, {\it {Fat Jets for a Light Higgs}},
  {\em Phys.Rev.Lett.} {\bf 104} (2010) 111801,
  [\href{http://xxx.lanl.gov/abs/0910.5472}{{\tt arXiv:0910.5472}}].

\bibitem{Thaler:2010tr}
J.~Thaler and K.~Van~Tilburg, {\it {Identifying Boosted Objects with
  N-subjettiness}},  {\em JHEP} {\bf 1103} (2011) 015,
  [\href{http://xxx.lanl.gov/abs/1011.2268}{{\tt arXiv:1011.2268}}].

\bibitem{Soper:2011cr}
D.~E. Soper and M.~Spannowsky, {\it {Finding physics signals with shower
  deconstruction}},  {\em Phys.Rev.} {\bf D84} (2011) 074002,
  [\href{http://xxx.lanl.gov/abs/1102.3480}{{\tt arXiv:1102.3480}}].

\bibitem{Englert:2011iz}
C.~Englert, T.~S. Roy, and M.~Spannowsky, {\it {Ditau jets in Higgs searches}},
   {\em Phys.Rev.} {\bf D84} (2011) 075026,
  [\href{http://xxx.lanl.gov/abs/1106.4545}{{\tt arXiv:1106.4545}}].

\bibitem{Thaler:2011gf}
J.~Thaler and K.~Van~Tilburg, {\it {Maximizing Boosted Top Identification by
  Minimizing N-subjettiness}},  {\em JHEP} {\bf 1202} (2012) 093,
  [\href{http://xxx.lanl.gov/abs/1108.2701}{{\tt arXiv:1108.2701}}].

\bibitem{Plehn:2011tg}
T.~Plehn and M.~Spannowsky, {\it {Top Tagging}},  {\em J.Phys.} {\bf G39}
  (2012) 083001, [\href{http://xxx.lanl.gov/abs/1112.4441}{{\tt
  arXiv:1112.4441}}].

\bibitem{Ellis:2012sd}
S.~D. Ellis, T.~S. Roy, and J.~Scholtz, {\it {Jets and Photons}},  {\em
  Phys.Rev.Lett.} {\bf 110} (2013), no.~12 122003,
  [\href{http://xxx.lanl.gov/abs/1210.1855}{{\tt arXiv:1210.1855}}].

\bibitem{Ellis:2012zp}
S.~D. Ellis, T.~S. Roy, and J.~Scholtz, {\it {Phenomenology of Photon-Jets}},
  {\em Phys.Rev.} {\bf D87} (2013) 014015,
  [\href{http://xxx.lanl.gov/abs/1210.3657}{{\tt arXiv:1210.3657}}].

\bibitem{Soper:2012pb}
D.~E. Soper and M.~Spannowsky, {\it {Finding top quarks with shower
  deconstruction}},  {\em Phys.Rev.} {\bf D87} (2013), no.~5 054012,
  [\href{http://xxx.lanl.gov/abs/1211.3140}{{\tt arXiv:1211.3140}}].

\bibitem{Ortiz:2014iza}
N.~G. Ortiz, J.~Ferrando, D.~Kar, and M.~Spannowsky, {\it {Reconstructing
  singly produced top partners in decays to $\mathbf{Wb}$}},
  \href{http://xxx.lanl.gov/abs/1403.7490}{{\tt arXiv:1403.7490}}.

\bibitem{Ellis:2012sn}
S.~D. Ellis, A.~Hornig, T.~S. Roy, D.~Krohn, and M.~D. Schwartz, {\it {Qjets: A
  Non-Deterministic Approach to Tree-Based Jet Substructure}},  {\em
  Phys.Rev.Lett.} {\bf 108} (2012) 182003,
  [\href{http://xxx.lanl.gov/abs/1201.1914}{{\tt arXiv:1201.1914}}].

\bibitem{Pedersen:2015knf}
K.~Pedersen and Z.~Sullivan, {\it {$\mu_x$ boosted-bottom-jet tagging and Z′
  boson searches}},  {\em Phys. Rev.} {\bf D93} (2016), no.~1 014014,
  [\href{http://xxx.lanl.gov/abs/1511.0599}{{\tt arXiv:1511.0599}}].

\bibitem{Pedersen:2015hka}
K.~Pedersen and Z.~Sullivan, {\it {Flavor tagging TeV jets for BSM and QCD}},
  in {\em {Meeting of the APS Division of Particles and Fields (DPF 2015) Ann
  Arbor, Michigan, USA, August 4-8, 2015}}, 2015.
\newblock \href{http://xxx.lanl.gov/abs/1509.0755}{{\tt arXiv:1509.0755}}.

\bibitem{Gallicchio:2010sw}
J.~Gallicchio and M.~D. Schwartz, {\it {Seeing in Color: Jet Superstructure}},
  {\em Phys.Rev.Lett.} {\bf 105} (2010) 022001,
  [\href{http://xxx.lanl.gov/abs/1001.5027}{{\tt arXiv:1001.5027}}].

\bibitem{Gallicchio:2011xc}
J.~Gallicchio and M.~D. Schwartz, {\it {Pure Samples of Quark and Gluon Jets at
  the LHC}},  {\em JHEP} {\bf 1110} (2011) 103,
  [\href{http://xxx.lanl.gov/abs/1104.1175}{{\tt arXiv:1104.1175}}].

\bibitem{Gallicchio:2011xq}
J.~Gallicchio and M.~D. Schwartz, {\it {Quark and Gluon Tagging at the LHC}},
  {\em Phys.Rev.Lett.} {\bf 107} (2011) 172001,
  [\href{http://xxx.lanl.gov/abs/1106.3076}{{\tt arXiv:1106.3076}}].

\bibitem{Gallicchio:2012ez}
J.~Gallicchio and M.~D. Schwartz, {\it {Quark and Gluon Jet Substructure}},
  {\em JHEP} {\bf 1304} (2013) 090,
  [\href{http://xxx.lanl.gov/abs/1211.7038}{{\tt arXiv:1211.7038}}].

\bibitem{Krohn:2012fg}
D.~Krohn, T.~Lin, M.~D. Schwartz, and W.~J. Waalewijn, {\it {Jet Charge at the
  LHC}},  \href{http://xxx.lanl.gov/abs/1209.2421}{{\tt arXiv:1209.2421}}.

\bibitem{Butterworth:2008sd}
J.~M. Butterworth, A.~R. Davison, M.~Rubin, and G.~P. Salam, {\it {Jet
  substructure as a new Higgs search channel at the LHC}},  {\em AIP
  Conf.Proc.} {\bf 1078} (2009) 189--191,
  [\href{http://xxx.lanl.gov/abs/0809.2530}{{\tt arXiv:0809.2530}}].

\bibitem{Butterworth:2008tr}
J.~M. Butterworth, A.~R. Davison, M.~Rubin, and G.~P. Salam, {\it {Jet
  substructure as a new Higgs search channel at the LHC}},
  \href{http://xxx.lanl.gov/abs/0810.0409}{{\tt arXiv:0810.0409}}.

\bibitem{Ellis:2009su}
S.~D. Ellis, C.~K. Vermilion, and J.~R. Walsh, {\it {Techniques for improved
  heavy particle searches with jet substructure}},  {\em Phys.Rev.} {\bf D80}
  (2009) 051501, [\href{http://xxx.lanl.gov/abs/0903.5081}{{\tt
  arXiv:0903.5081}}].

\bibitem{Ellis:2009me}
S.~D. Ellis, C.~K. Vermilion, and J.~R. Walsh, {\it {Recombination Algorithms
  and Jet Substructure: Pruning as a Tool for Heavy Particle Searches}},  {\em
  Phys.Rev.} {\bf D81} (2010) 094023,
  [\href{http://xxx.lanl.gov/abs/0912.0033}{{\tt arXiv:0912.0033}}].

\bibitem{Krohn:2009th}
D.~Krohn, J.~Thaler, and L.-T. Wang, {\it {Jet Trimming}},  {\em JHEP} {\bf
  1002} (2010) 084, [\href{http://xxx.lanl.gov/abs/0912.1342}{{\tt
  arXiv:0912.1342}}].

\bibitem{Soyez:2012hv}
G.~Soyez, G.~P. Salam, J.~Kim, S.~Dutta, and M.~Cacciari, {\it {Pileup
  subtraction for jet shapes}},  {\em Phys.Rev.Lett.} {\bf 110} (2013), no.~16
  162001, [\href{http://xxx.lanl.gov/abs/1211.2811}{{\tt arXiv:1211.2811}}].

\bibitem{Cacciari:2014jta}
M.~Cacciari, G.~P. Salam, and G.~Soyez, {\it {On the use of charged-track
  information to subtract neutral pileup}},
  \href{http://xxx.lanl.gov/abs/1404.7353}{{\tt arXiv:1404.7353}}.

\bibitem{Cacciari:2014gra}
M.~Cacciari, G.~P. Salam, and G.~Soyez, {\it {SoftKiller, a particle-level
  pileup removal method}},  \href{http://xxx.lanl.gov/abs/1407.0408}{{\tt
  arXiv:1407.0408}}.

\bibitem{Bertolini:2014bba}
D.~Bertolini, P.~Harris, M.~Low, and N.~Tran, {\it {Pileup Per Particle
  Identification}},  \href{http://xxx.lanl.gov/abs/1407.6013}{{\tt
  arXiv:1407.6013}}.

\bibitem{Aad:2015owa}
{\bf ATLAS} Collaboration, G.~Aad et~al., {\it {Search for high-mass diboson
  resonances with boson-tagged jets in proton-proton collisions at $ \sqrt{s}=8
  $ TeV with the ATLAS detector}},  {\em JHEP} {\bf 12} (2015) 055,
  [\href{http://xxx.lanl.gov/abs/1506.0096}{{\tt arXiv:1506.0096}}].

\bibitem{Brehmer:2015dan}
J.~Brehmer et~al., {\it {The Diboson Excess: Experimental Situation and
  Classification of Explanations; A Les Houches Pre-Proceeding}},
  \href{http://xxx.lanl.gov/abs/1512.0435}{{\tt arXiv:1512.0435}}.

\bibitem{Goncalves:2015yua}
D.~Gonçalves, F.~Krauss, and M.~Spannowsky, {\it {Augmenting the diboson excess
  for the LHC Run II}},  {\em Phys. Rev.} {\bf D92} (2015), no.~5 053010,
  [\href{http://xxx.lanl.gov/abs/1508.0416}{{\tt arXiv:1508.0416}}].

\bibitem{Aaltonen:2014mdq}
{\bf CDF} Collaboration, T.~Aaltonen et~al., {\it {Invariant-mass distribution
  of jet pairs produced in association with a $W$ boson in $p \bar{p}$
  collisions at $\sqrt{s}=1.96$ TeV using the full CDF Run II data set}},  {\em
  Phys. Rev.} {\bf D89} (2014), no.~9 092001,
  [\href{http://xxx.lanl.gov/abs/1402.7044}{{\tt arXiv:1402.7044}}].

\bibitem{Aaltonen:2011mk}
{\bf CDF} Collaboration, T.~Aaltonen et~al., {\it {Invariant Mass Distribution
  of Jet Pairs Produced in Association with a $W$ boson in $p \bar{p}$
  Collisions at $\sqrt{s}= 1.96$ TeV}},  {\em Phys. Rev. Lett.} {\bf 106}
  (2011) 171801, [\href{http://xxx.lanl.gov/abs/1104.0699}{{\tt
  arXiv:1104.0699}}].

\bibitem{Khachatryan:2014hpa}
{\bf CMS} Collaboration, V.~Khachatryan et~al., {\it {Search for massive
  resonances in dijet systems containing jets tagged as W or Z boson decays in
  pp collisions at $ \sqrt{s} $ = 8 TeV}},  {\em JHEP} {\bf 08} (2014) 173,
  [\href{http://xxx.lanl.gov/abs/1405.1994}{{\tt arXiv:1405.1994}}].

\bibitem{CMS-PAS-EXO-15-002}
{\bf CMS Collaboration} Collaboration, {\it {Search for massive resonances
  decaying into pairs of boosted W and Z bosons at $\sqrt{s}$ = 13 TeV}},
  Tech. Rep. CMS-PAS-EXO-15-002, CERN, Geneva, 2015.

\bibitem{ATLAS-CONF-2015-073}
{\it {Search for resonances with boson-tagged jets in 3.2 fb?1 of p p
  collisions at ? s = 13 TeV collected with the ATLAS detector}},  Tech. Rep.
  ATLAS-CONF-2015-073, CERN, Geneva, Dec, 2015.

\bibitem{Kribs:2009yh}
G.~D. Kribs, A.~Martin, T.~S. Roy, and M.~Spannowsky, {\it {Discovering the
  Higgs Boson in New Physics Events using Jet Substructure}},  {\em Phys.Rev.}
  {\bf D81} (2010) 111501, [\href{http://xxx.lanl.gov/abs/0912.4731}{{\tt
  arXiv:0912.4731}}].

\bibitem{Kribs:2010hp}
G.~D. Kribs, A.~Martin, T.~S. Roy, and M.~Spannowsky, {\it {Discovering Higgs
  Bosons of the MSSM using Jet Substructure}},  {\em Phys.Rev.} {\bf D82}
  (2010) 095012, [\href{http://xxx.lanl.gov/abs/1006.1656}{{\tt
  arXiv:1006.1656}}].

\bibitem{Kribs:2010ii}
G.~D. Kribs, A.~Martin, and T.~S. Roy, {\it {Higgs boson discovery through
  top-partners decays using jet substructure}},  {\em Phys.Rev.} {\bf D84}
  (2011) 095024, [\href{http://xxx.lanl.gov/abs/1012.2866}{{\tt
  arXiv:1012.2866}}].

\bibitem{Ellis:2014eya}
S.~D. Ellis, A.~Hornig, D.~Krohn, and T.~S. Roy, {\it {On Statistical Aspects
  of Qjets}},  {\em JHEP} {\bf 01} (2015) 022,
  [\href{http://xxx.lanl.gov/abs/1409.6785}{{\tt arXiv:1409.6785}}].

\bibitem{Altheimer:2013yza}
A.~Altheimer, A.~Arce, L.~Asquith, J.~Backus~Mayes, E.~Bergeaas~Kuutmann,
  et~al., {\it {Boosted objects and jet substructure at the LHC. Report of
  BOOST2012, held at IFIC Valencia, 23rd-27th of July 2012}},  {\em
  Eur.Phys.J.} {\bf C74} (2014) 2792,
  [\href{http://xxx.lanl.gov/abs/1311.2708}{{\tt arXiv:1311.2708}}].

\bibitem{Dokshitzer:1997in}
Y.~L. Dokshitzer, G.~D. Leder, S.~Moretti, and B.~R. Webber, {\it {Better jet
  clustering algorithms}},  {\em JHEP} {\bf 08} (1997) 001,
  [\href{http://xxx.lanl.gov/abs/hep-ph/9707323}{{\tt hep-ph/9707323}}].

\bibitem{Wobisch:1998wt}
M.~Wobisch and T.~Wengler, {\it {Hadronization corrections to jet
  cross-sections in deep inelastic scattering}},  in {\em {Monte Carlo
  generators for HERA physics. Proceedings, Workshop, Hamburg, Germany,
  1998-1999}}, 1998.
\newblock \href{http://xxx.lanl.gov/abs/hep-ph/9907280}{{\tt hep-ph/9907280}}.

\bibitem{Wobisch:2000dk}
M.~Wobisch, {\it {Measurement and QCD analysis of jet cross-sections in deep
  inelastic positron proton collisions at s**(1/2) = 300-GeV}}, .

\bibitem{Ellis:1993tq}
S.~D. Ellis and D.~E. Soper, {\it {Successive combination jet algorithm for
  hadron collisions}},  {\em Phys. Rev.} {\bf D48} (1993) 3160--3166,
  [\href{http://xxx.lanl.gov/abs/hep-ph/9305266}{{\tt hep-ph/9305266}}].

\bibitem{Catani:1993hr}
S.~Catani, Y.~L. Dokshitzer, M.~H. Seymour, and B.~R. Webber, {\it
  {Longitudinally invariant $K_t$ clustering algorithms for hadron hadron
  collisions}},  {\em Nucl. Phys.} {\bf B406} (1993) 187--224.

\bibitem{Brodsky:1976mg}
S.~J. Brodsky and J.~F. Gunion, {\it {Hadron Multiplicity in Color Gauge Theory
  Models}},  {\em Phys. Rev. Lett.} {\bf 37} (1976) 402--405.

\bibitem{Gaffney:1984yd}
J.~B. Gaffney and A.~H. Mueller, {\it {Alpha (Q**2) Corrections to Particle
  Multiplicity Ratios in Gluon and Quark Jets}},  {\em Nucl. Phys.} {\bf B250}
  (1985) 109.

\bibitem{Aad:2014gea}
{\bf ATLAS} Collaboration, G.~Aad et~al., {\it {Light-quark and gluon jet
  discrimination in $pp$ collisions at $\sqrt{s}=7\mathrm {\ TeV}$ with the
  ATLAS detector}},  {\em Eur. Phys. J.} {\bf C74} (2014), no.~8 3023,
  [\href{http://xxx.lanl.gov/abs/1405.6583}{{\tt arXiv:1405.6583}}].

\bibitem{Sjostrand:2006za}
T.~Sjostrand, S.~Mrenna, and P.~Z. Skands, {\it {PYTHIA 6.4 Physics and
  Manual}},  {\em JHEP} {\bf 05} (2006) 026,
  [\href{http://xxx.lanl.gov/abs/hep-ph/0603175}{{\tt hep-ph/0603175}}].

\bibitem{Sjostrand:2007gs}
T.~Sjostrand, S.~Mrenna, and P.~Z. Skands, {\it {A Brief Introduction to PYTHIA
  8.1}},  {\em Comput. Phys. Commun.} {\bf 178} (2008) 852--867,
  [\href{http://xxx.lanl.gov/abs/0710.3820}{{\tt arXiv:0710.3820}}].

\bibitem{deFavereau:2013fsa}
{\bf DELPHES 3} Collaboration, J.~de~Favereau, C.~Delaere, P.~Demin,
  A.~Giammanco, V.~Lemaître, A.~Mertens, and M.~Selvaggi, {\it {DELPHES 3, A
  modular framework for fast simulation of a generic collider experiment}},
  {\em JHEP} {\bf 02} (2014) 057,
  [\href{http://xxx.lanl.gov/abs/1307.6346}{{\tt arXiv:1307.6346}}].

\bibitem{Cacciari:2005hq}
M.~Cacciari and G.~P. Salam, {\it {Dispelling the $N^{3}$ myth for the $k_t$
  jet-finder}},  {\em Phys. Lett.} {\bf B641} (2006) 57--61,
  [\href{http://xxx.lanl.gov/abs/hep-ph/0512210}{{\tt hep-ph/0512210}}].

\bibitem{Cacciari:2011ma}
M.~Cacciari, G.~P. Salam, and G.~Soyez, {\it {FastJet User Manual}},  {\em Eur.
  Phys. J.} {\bf C72} (2012) 1896,
  [\href{http://xxx.lanl.gov/abs/1111.6097}{{\tt arXiv:1111.6097}}].

\end{thebibliography}\endgroup

\end{document}